\documentclass[letterpaper,journal]{IEEEtran}
\usepackage{amsmath,amssymb,amsfonts,amsthm}
\usepackage{array}
\usepackage{textcomp}
\usepackage{stfloats}
\usepackage{url}
\usepackage{graphicx}
\usepackage{cite}
\usepackage{booktabs}
\usepackage{multirow}
\usepackage{microtype}
\usepackage{algorithm}
\usepackage{algpseudocode}

\newtheorem{proposition}{Proposition}
\newtheorem{corollary}{Corollary}
\newtheorem{remark}{Remark}

\newcommand{\E}{\mathbb{E}}
\newcommand{\Prob}{\mathbb{P}}
\newcommand{\Ind}{\mathbb{I}}
\newcommand{\R}{\mathbb{R}}
\newcommand{\Cset}{\mathbb{C}}
\newcommand{\calC}{\mathcal{C}}

\newcommand{\calY}{\mathcal{Y}}
\newcommand{\Zsig}{Z_{\mathrm{sig}}}
\newcommand{\Zsub}{Z_{\mathrm{sub}}}
\newcommand{\Zbs}{Z_{\mathrm{bs}}}
\newcommand{\ESS}{\mathrm{ESS}}
\newcommand{\ASS}{\mathrm{ASS}}
\newcommand{\MAE}{\mathrm{MAE}}

\begin{document}

\title{PhyCal: Physics-Informed Unlabeled Conformal Calibration for Automatic Modulation Classification Under Channel Shift}

\author{Yezhuo~Zhang,
        Zinan~Zhou,
        Ruixiang~Zhang,
        and~Xuanpeng~Li% <-this % stops a space
\thanks{Manuscript received [Month Day, Year]; revised [Month Day, Year]; accepted [Month Day, Year]. This work was supported by the Fundamental Research Funds for the Central Universities (Grant Nos. 2242026K30005 and 2242026K30024). (Corresponding author: Xuanpeng Li.)}% <-this % stops a space
\thanks{The authors are with the School of Instrument Science and Engineering, State Key Laboratory of Comprehensive PNT Network and Equipment Technology, and Key Laboratory of Micro-Inertial Instrument and Advanced Navigation Technology, Ministry of Education, Southeast University, Nanjing 210096, China (e-mail: zhang\_yezhuo@seu.edu.cn; zhouzinan919@seu.edu.cn; zhang\_ruixiang@seu.edu.cn; li\_xuanpeng@seu.edu.cn).}}

% The paper headers
\markboth{IEEE Transactions on Cognitive Communications and Networking,~Vol.~XX, No.~X, Month~Year}%
{Zhang \MakeLowercase{\textit{et al.}}: PhyCal: Physics-Informed Unlabeled Conformal Calibration for AMC}

% Per IEEE instructions, \IEEEpubid is not required at initial submission for
% Transactions/Journals (added by IEEE production). If enabled later, also call
% \IEEEpubidadjcol in the second column of page 1:
% \IEEEpubid{0000--0000/00\$00.00~\copyright~2026 IEEE}

\maketitle

\begin{abstract}
Deep automatic modulation classification (AMC) achieves high accuracy in matched channels, yet deployed receivers lack reliable confidence calibration once wireless channels shift: existing methods assume stationary links or labeled target waveforms unavailable in live over-the-air (OTA) operation.
We propose PhyCal, a classifier-agnostic, zero-label conformal calibration framework wrapping any frozen pre-trained AMC model to output calibrated prediction sets.
Grounded in baseband physics, PhyCal diagnoses target signal quality from received waveforms without consulting the classifier, simulates channel degradations on labeled source data to learn conditional true-class rank distributions, and transports source residuals via a convex channel mixture at full effective sample size.
Across three AMC architectures, PhyCal outperforms all 12 unlabeled baselines on synthetic signal-to-noise ratio (SNR) drops, real OTA distance shifts, and overall four-scenario real 2.4~GHz multipath reliability, matches target-labeled calibration, and equalizes coverage across unknown SNRs.
At 99\% target coverage, it cuts candidate-hypothesis blind demodulation workload by up to 61\%, scales to a single-pass edge variant adding 0.3\% multiply--accumulate overhead ($9.2\text{--}11.7\times$ single-thread CPU speedup over multi-pass PhyCal), and stays robust under stationary links, small buffers, and skewed class priors.
Code and artifacts are released at \url{https://github.com/ZhangYezhuo/PhyCal}.
\end{abstract}

\begin{IEEEkeywords}
Automatic modulation classification, conformal prediction, uncertainty quantification, unlabeled calibration, wireless channel shift.
\end{IEEEkeywords}

% =============================================================================
\section{Introduction}
\label{sec:intro}
% =============================================================================

\IEEEPARstart{A}{utomatic} modulation classification (AMC) enables cognitive radios, spectrum monitors, and multi-standard receivers to identify an incoming signal's modulation prior to demodulation~\cite{nandi1998algorithms,oshea2018overtheair,sathyanarayanan2023rml22}.
Modern deep AMC networks achieve strong accuracy when training and deployment channels match~\cite{he2016resnet,oshea2018overtheair,xu2020mcldnn,trabelsi2018deep,zhang2026cspmnet}, yet practical links face abrupt drops in signal-to-noise ratio (SNR), multipath fading, carrier frequency offset (CFO), and over-the-air (OTA) distance or radio-frequency (RF) front-end drift~\cite{utrilla2020migou,belousov2026multipath}.
Under such shifts, deep classifiers become silently overconfident: a single unflagged misclassification routes an in-phase/quadrature (I/Q) frame to an incompatible demodulator, causing unrecoverable packet loss or wasted baseband processing.

Despite this vulnerability, AMC research has prioritized top-1 accuracy over \emph{confidence calibration under channel shift}.
Recent uncertainty quantification (UQ) studies explore Bayesian neural networks~\cite{luu2024bayesian}, deep ensembles~\cite{yang2026uq}, and conformal prediction (CP)~\cite{cohen2023wirelesscp,conformalshield2025}, yet Bayesian and ensemble methods lack explicit coverage control under distribution shift, whereas existing wireless CP schemes assume stationary channels or \emph{labeled} target calibration sets unavailable in live or non-cooperative links.

Conformal prediction~\cite{vovk2005algorithmic,lei2018distribution,sadinle2019least} offers a principled reliability layer for black-box AMC models: wrapping \emph{any} pre-trained classifier, it outputs a calibrated \emph{prediction set} of candidate modulations covering the true class at a target confidence such as 90\% or 99\%.
In a parallel multi-hypothesis blind demodulator~\cite{cohen2023wirelesscp}, high-SNR frames contract to a singleton while degraded frames retain only plausible candidates, avoiding both exhaustive all-class decoding and silent top-1 failures under high-reliability, low-latency (e.g., ultra-reliable low-latency communications [URLLC]) budgets.

Adapting domain-agnostic shift-aware CP from machine learning---such as feature importance weighting~\cite{tibshirani2019wcp,park2020cpcs}, doubly robust feature calibration~\cite{qiu2023adaptive,yang2024doubly}, test-time score or optimal-transport alignment~\cite{correia2025otcp,kasa2025ecp,yilmaz2022qtc}, and Mixup or pseudo-label calibration~\cite{hu2024pseudocal,angelman2025sfcp}---proves brittle on complex baseband I/Q waveforms.
Relying solely on neural features or shifted softmax scores without RF propagation physics leads to severe under-coverage or vacuous sets under steep SNR drops and multipath distortion (Sections~\ref{sec:related_work} and~\ref{sec:exp_core}).

To address this from baseband physics, we propose \textbf{PhyCal}, a classifier-agnostic framework for zero-label conformal calibration under channel shift, grounded in \emph{carrier-phase rotational symmetry} ($U(1)$ invariance), \emph{one-way classifier-free degradation}, and \emph{noise monotonicity} (Section~\ref{sec:method}).
Our contributions are threefold:
\begin{itemize}
    \item \textbf{Physics-Informed Unlabeled CP and Theory}: PhyCal wraps any frozen AMC model, replacing unstable feature reweighting with classifier-free baseband descriptors and uniform-weight channel-mixture transport. We prove global mixture optimality at maximal source effective sample size (ESS), doubly robust error bounds with identifiability limits, and approximate coverage control.
    \item \textbf{Edge Efficiency and High-Reliability Savings}: Single-pass \textbf{PhyCal-Lite} adds only $0.3\%$ multiply--accumulate (MAC) overhead ($9.2\text{--}11.7\times$ CPU speedup over multi-pass PhyCal) while outperforming the full variant on unadapted models. At 99\% coverage, PhyCal avoids full-set collapse and cuts candidate-hypothesis demodulation workload by up to $61\%$.
    \item \textbf{Cross-Architecture Validation and Robustness}: Across different AMC architectures (unadapted and adapted), PhyCal outperforms all unlabeled baselines on synthetic SNR drops, real OTA distance shifts, and overall multipath reliability. It matches target-labeled calibration, cuts per-SNR spread $5\text{--}7\times$, and stays robust under stationary links, small buffers, and skewed priors.
\end{itemize}

The remainder of this paper covers formulation and related work (Section~\ref{sec:problem}), PhyCal and theory (Section~\ref{sec:method}), experiments (Section~\ref{sec:experiments}), and conclusions (Section~\ref{sec:conclusion}).

% =============================================================================
\section{Problem Formulation and Related Work}
\label{sec:problem}
% =============================================================================

\subsection{Unlabeled Calibration Under Wireless Channel Shift}
\label{sec:prob_setup}

\begin{table}[!t]
\centering
\caption{Summary of Key Notation}
\label{tab:notation}
\footnotesize
\renewcommand{\arraystretch}{0.94}
\setlength{\tabcolsep}{3.2pt}
\begin{tabular}{@{}l>{\raggedright\arraybackslash}p{5.85cm}@{}}
\toprule
\textbf{Symbol} & \textbf{Meaning and Role} \\
\midrule
\multicolumn{2}{@{}l}{\textit{Operational Scenario and Receiver Streams (Section~\ref{sec:prob_setup})}} \\
$z\!\in\!\Cset^L$, $x\!\in\!\R^{2\times L}$ & Unit-RMS complex baseband frame and real I/Q tensor \\
$f(x)$, $p_f(y\mid x)$ & Strictly frozen pre-trained AMC model and probabilities \\
$C_p$ ($N_{\mathrm{base}}$) & Offline labeled source pool of $N_{\mathrm{base}}$ base frames ($P_S$) \\
$A$ ($N_A$), $U$ ($N_U$) & Unlabeled $P_T$ adaptation and calibration windows \\
$T$ ($N_T$) & Subsequent online $P_T$ reception stream (true $y_n^T$ unknown) \\
$\calC_\alpha(x)$, $\ASS$ & Prediction set at miscoverage $\alpha$ and average set size \\
\midrule
\multicolumn{2}{@{}l}{\textit{PhyCal Calibration Framework (Section~\ref{sec:method})}} \\
$\Zsub$, $\Zbs$, $\Zsig$ & 9-D Toeplitz, 6-D blind-spot, and 15-D signal descriptors \\
$\zeta(x)$ & Classifier-free descriptor ($\Zsig(x)$, or $\Zsub(x)$ in PhyCal-Lite) \\
$\tilde{q}(r\mid x)$ & Calibrated 5-fold out-of-fold conditional rank posterior \\
$\Delta_{\mathrm{snr}}$, $\hat{\pi}\!\in\!\Delta^{V-1}$ & Unlabeled SNR shift and mixture over $V = 8$ channel views \\
$w_{v,i}^* = \hat{\pi}_v / N_{\mathrm{base}}$ & Uniform source weights (maximizing $\ESS$; Prop.~\ref{prop:em_ess}) \\
$\widetilde{F}_T(t)$, $\hat{\tau}_\alpha$ & Calibrated target score CDF and $(1{-}\alpha)$ threshold \\
\bottomrule
\end{tabular}
\end{table}

Consider a receiver processing unit-root-mean-square (unit-RMS) baseband frames $z = z_I + j z_Q \in \Cset^L$ ($j = \sqrt{-1}$, $\frac{1}{L}\sum_{n=0}^{L-1} |z[n]|^2 = 1$), represented as I/Q tensors $x \in \R^{2\times L}$ carrying unknown modulation $Y \in \calY = \{1,\dots,K\}$.
Before deployment, the receiver holds a frozen AMC classifier $f:\R^{2\times L}\to \Delta^{K-1}$ with outputs $p_f(y\mid x)$, plus an offline labeled source pool $C_p = \{(x_i^S, y_i^S)\}_{i=1}^{N_{\mathrm{base}}}$ recorded under a baseline channel $P_S$ (e.g., high-SNR, short-range OTA, or line-of-sight [LoS]).

Under a shifted channel $P_T \neq P_S$, the receiver observes only unlabeled waveforms (\emph{zero labels}, \emph{zero SNR metadata}) across three disjoint windows from $P_T$ (Table~\ref{tab:notation}):
(i)~an \emph{adaptation window} $A = \{x_a^A\}_{a=1}^{N_A}$ to diagnose channel conditions and fit auxiliary modules;
(ii)~a fresh \emph{calibration window} $U = \{x_m^U\}_{m=1}^{N_U}$ to set the conformal threshold (kept disjoint from $A$ to prevent in-sample bias); and
(iii)~the \emph{online stream} $T = \{x_n^T\}_{n=1}^{N_T}$ with unknown labels $y_n^T$.

For target miscoverage $\alpha \in (0,1)$ (e.g., $1-\alpha = 90\%$ or $99\%$ for high-reliability operation), an unlabeled calibrator uses only $(f, C_p, A, U)$ to construct a nonconformity score $\tilde{s}(x,y)\in\R$ (smaller denotes higher plausibility, built from $s(x,y) = -\log p_f(y\mid x) \ge 0$~\cite{sadinle2019least}) and threshold $\hat{\tau}_\alpha$, outputting for each $x \in T$:
\begin{equation}
    \calC_\alpha(x) = \left\{ y \in \calY : \tilde{s}(x,y) \le \hat{\tau}_\alpha \right\}.
    \label{eq:conf_set}
\end{equation}
A reliable receiver requires $\calC_\alpha$ to satisfy four objectives under $P_T$:
\begin{enumerate}
    \item \emph{Marginal reliability}: $\Prob_{P_T}\{Y \in \calC_\alpha(X)\}$ tracks $1-\alpha$, avoiding both silent under-coverage and vacuous sets.
    \item \emph{Compactness and hypothesis reduction}: Minimizing average set size $\ASS = \E_{P_T}[|\calC_\alpha(X)|]$ reduces candidate modulations sent to a parallel multi-hypothesis demodulator vs.\ exhaustive $K$-class decoding, tracked by the first-order hypothesis-reduction proxy
    \begin{equation}
        \mathrm{Saving} = \left(1 - \frac{\ASS}{K}\right)\times 100\%,
        \label{eq:demod_saving}
    \end{equation}
    which assumes comparable per-hypothesis demodulation cost (not measured demodulator latency or block error rate) and is valid only when nominal coverage $1-\alpha$ is met.
    \item \emph{SNR-conditional uniformity}: $\Prob_{P_T}\{Y\in\calC_\alpha(X)\mid \mathrm{SNR}\in g\}$ stays flat across unobserved SNR strata $g$.
    \item \emph{Hard-class protection}: $\min_{y\in\calY} \Prob_{P_T}\{Y\in\calC_\alpha(X)\mid Y=y\}$ stays high so dense constellations (e.g., 16-QAM, QPSK) and minority classes are not sacrificed to satisfy a marginal average.
\end{enumerate}

\subsection{Related Work}
\label{sec:related_work}

\subsubsection{Calibration, Uncertainty Quantification, and Conformal Prediction}
Post-hoc and UQ methods such as temperature scaling~\cite{guo2017calibration}, Monte Carlo dropout~\cite{gal2016dropout}, and deep ensembles~\cite{lakshminarayanan2017ensembles} improve calibration yet lack finite-sample coverage guarantees.
Conformal prediction~\cite{vovk2005algorithmic,lei2018distribution} wraps any classifier to guarantee $\Prob\{Y\in\calC_\alpha(X)\}\ge 1-\alpha$ under exchangeability: least ambiguous set-valued classifiers (LAC)~\cite{sadinle2019least} use $s_{\mathrm{LAC}}(x,y)=-\log p_f(y\mid x)$ for minimal sets, adaptive/regularized prediction sets (APS/RAPS)~\cite{romano2020aps,angelopoulos2021raps} sum sorted probabilities for conditional adaptivity, and conformalized quantile regression (CQR)~\cite{koenker1978regression,romano2019cqr} and distributional CP (DCP)~\cite{chernozhukov2021dcp} form pivotal scores from conditional quantiles or cumulative distribution functions (CDFs).
Channel shift breaks exchangeability and invalidates split CP.

\subsubsection{Unlabeled and Shift-Aware Calibration}
Domain-agnostic shift calibrators span four families:
(i)~feature importance weighting---weighted CP (WCP)~\cite{tibshirani2019wcp}, calibrated prediction with covariate shift (CPCS)~\cite{park2020cpcs}, and transferable calibration (TransCal)~\cite{wang2020transcal}---via estimated density ratios $dP_T/dP_S$;
(ii)~doubly robust (DR) shift calibration~\cite{qiu2023adaptive,yang2024doubly}, rooted in semiparametric inference~\cite{robins1994estimation}, double/debiased machine learning (DML)~\cite{chernozhukov2018dml}, and prediction-powered inference (PPI)~\cite{angelopoulos2023ppi};
(iii)~test-time alignment including quantile test-time calibration (QTC)~\cite{yilmaz2022qtc}, entropy-scaled CP (ECP)~\cite{kasa2025ecp}, and optimal-transport CP (OT-CP)~\cite{correia2025otcp}; and
(iv)~synthetic interpolation or self-training such as Mixup pseudo-calibration (PseudoCal)~\cite{hu2024pseudocal} and source-free pseudo-label CP (SFCP)~\cite{angelman2025sfcp}.
On I/Q waveforms, these fail via three mechanisms (Section~\ref{sec:exp_core}): steep SNR drops collapse density-ratio $\ESS$; low-SNR overconfidence corrupts score quantiles and pseudo-labels; and linear I/Q Mixup creates unphysical co-channel interference.

\subsubsection{Domain Adaptation, UQ, and Calibration in AMC}
Deep AMC excels on matched channels~\cite{oshea2018overtheair,xu2020mcldnn,trabelsi2018deep,sathyanarayanan2023rml22,zhang2026cspmnet} but degrades under SNR drops, multipath, and hardware drift~\cite{utrilla2020migou,belousov2026multipath}.
When retraining is allowed, domain-adversarial neural networks (DANN)~\cite{ganin2016dann}, conditional adversarial domain adaptation (CDAN)~\cite{long2018cdan}, and minimum class confusion (MCC)~\cite{jin2020mcc} raise accuracy yet still lack shift-aware coverage control.
Conversely, AMC UQ via Bayesian networks~\cite{luu2024bayesian}, ensembles~\cite{yang2026uq}, or split CP for wireless transceivers~\cite{cohen2023wirelesscp} and adversarial detection~\cite{conformalshield2025} either degrade under channel shift or require stationary channels or target labels, motivating a strictly post-hoc, physics-grounded, zero-label CP wrapper for both unadapted and adapted frozen models.

% =============================================================================
\section{The PhyCal Framework and Theoretical Analysis}
\label{sec:method}
% =============================================================================

\begin{figure*}[!t]
    \centering
    \includegraphics[width=1.00\textwidth]{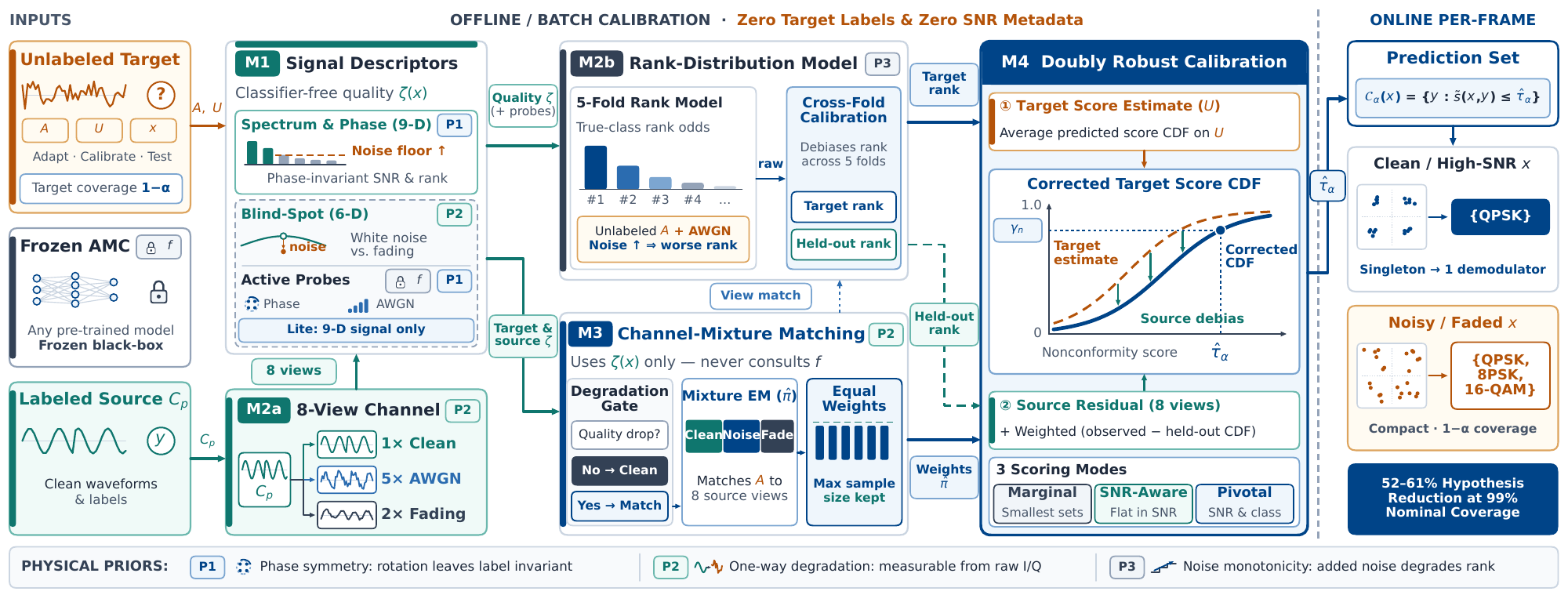}
    \caption{Overview of \textbf{PhyCal}: M1 extracts classifier-free I/Q descriptors; M2a generates eight synthetic source channel views and M2b fits a cross-fold rank-distribution model using $A$; M3 estimates a convex channel-view mixture; and M4 combines target CDF predictions on $U$ with mixture-weighted OOF source residuals to set $\hat{\tau}_\alpha$ without target labels or SNR metadata.}
    \label{fig:framework}
\end{figure*}

\subsection{Wireless Physical Priors and Pipeline Overview}
\label{sec:method_overview}

Setting $\hat{\tau}_\alpha$ in~\eqref{eq:conf_set} without target labels amounts to estimating the unknown target score CDF $F_T(t) = \Prob_{P_T}\{\tilde{s}(X,Y) \le t\}$.
Under shift, source thresholds fail ($F_S \neq F_T$) while plug-in target predictions inherit $f$'s miscalibration.
PhyCal pairs \emph{classifier-free baseband descriptors} with a \emph{synthetic RF channel suite} (Fig.~\ref{fig:framework}), anchored in three physical priors:
\begin{itemize}
    \item \textbf{P1 (Carrier-Phase Symmetry)}: Oscillator mismatch rotates constellations via $z \mapsto e^{j\phi}z$ ($\phi \in [0,2\pi)$), leaving modulation $Y$ and intrinsic quality invariant; sensitivity of $f$ across $\phi$ thus exposes instability.
    \item \textbf{P2 (One-Way Degradation \& Observability)}: Additive white Gaussian noise (AWGN), multipath fading, and CFO degrade waveforms strictly forward independently of $f$, leaving noise-floor and correlation cues measurable directly from raw I/Q.
    \item \textbf{P3 (Noise Monotonicity)}: Adding independent synthetic AWGN forms a Blackwell-garbled experiment~\cite{blackwell1953equivalent} degrading Bayes-optimal rank identifiability, motivating label-free monotonicity regularization on $A$.
\end{itemize}

Moving channel diagnosis into a low-dimensional classifier-free space bypasses both overconfident probabilities and high-variance neural density ratios via four modules (Fig.~\ref{fig:framework}):
M1 extracts $U(1)$-invariant subspace features and blind-spot residuals $\zeta(x)$ (P1--P2);
M2 passes $C_p$ through $V = 8$ simulated forward \emph{channel views} (M2a) to learn conditional true-class rank degradation with noise monotonicity (P3) on $A$ (M2b);
M3 applies a one-way degradation gate (P2) and matches $A$ to a convex view mixture in $\zeta$-space with uniform frame weights to preserve $\ESS$; and
M4 combines target predictions on fresh $U$ with channel-matched out-of-fold (OOF) source residuals to estimate $F_T(t)$ and invert it for $\hat{\tau}_\alpha$.

\subsection{Classifier-Free Signal Descriptors and Deployment Configurations (Module~1)}
\label{sec:module1}

Module~1 leaves every waveform passed to $f$ untouched and extracts physical descriptors directly from unit-RMS frames $z[n]$ ($x \leftrightarrow z$) without consulting $f$ (closed forms in Table~\ref{tab:descriptors}).

\begin{table}[!t]
\centering
\caption{Classifier-Free Baseband Signal Descriptors in Module~1 ($\Zsig = [\Zsub;\,\Zbs]\in\R^{15}$)}
\label{tab:descriptors}
\footnotesize
\renewcommand{\arraystretch}{0.95}
\setlength{\tabcolsep}{3.2pt}
\begin{tabular}{@{}l>{\raggedright\arraybackslash}p{5.55cm}@{}}
\toprule
\textbf{Descriptor} & \textbf{Closed-Form Definition} \\
\midrule
\multicolumn{2}{@{}l}{\textit{Autocorrelation Subspace Block $\Zsub$ (9-D, Strictly $U(1)$-Invariant)}} \\
Noise-tail floor & $\lambda_{\mathrm{low}} / \bar{\lambda}$, with $\lambda_{\mathrm{low}} = \frac{1}{2}(\lambda_7+\lambda_8)$ \\
Log blind-SNR proxy & $\log\big((\lambda_{\mathrm{high}} + \epsilon)/(\lambda_{\mathrm{low}} + \epsilon)\big)$ \\
Log condition ratio & $\log\big((\lambda_1 + \epsilon)/(\lambda_8 + \epsilon)\big)$ \\
Spectral effective rank & $\frac{1}{8}\exp\!\big(-\sum_{m=1}^{8} e_m \log e_m\big)$ \\
DFT floor quantile $q_{10}$ & $L \cdot \mathrm{Quantile}_{0.10}(\hat{S}_z)$ \\
DFT floor quantile $q_{25}$ & $L \cdot \mathrm{Quantile}_{0.25}(\hat{S}_z)$ \\
Lag-1 phase-diff SD & $\mathrm{std}(\Delta\phi[n])$, $\Delta\phi[n]=\arg(z[n]z^*[n-1])$ \\
Phase-diff coherence & $\big|\frac{1}{L-1}\sum_{n=1}^{L-1} e^{j\Delta\phi[n]}\big|$ \\
Envelope IQR ratio & $\big(Q_{0.75}(|z|) - Q_{0.25}(|z|)\big)/Q_{0.50}(|z|)$ \\
\midrule
\multicolumn{2}{@{}l}{\textit{Blind-Spot Residual Block $\Zbs$ (6-D, 15.0K $J$-Invariant Denoiser)}} \\
Blind-spot log SNR & $\log\big((P_{\hat{z}} + \epsilon')/(P_w + \epsilon')\big)$ \\
Residual power ratio & $P_w / P_z$ \\
Lag-1 resid.\ whiteness & $\big|\overline{w[n]w^*[n-1]}\big| / P_w$ \\
Lag-2 resid.\ whiteness & $\big|\overline{w[n]w^*[n-2]}\big| / P_w$ \\
Resid.\ spec.\ flatness & $\exp\!\big(\overline{\log |W[k]|^2}\big) / \overline{|W[k]|^2}$ \\
Signal--pred.\ corr. & $\overline{\mathrm{Re}(z\hat{z}^*)} / \sqrt{P_z P_{\hat{z}}}$ \\
\bottomrule
\end{tabular}
\vspace{1pt}
\parbox{\columnwidth}{\scriptsize $\lambda_1 \ge \cdots \ge \lambda_8 \ge 0$: eigenvalues of $8\times 8$ Hermitian positive-semidefinite Toeplitz $\mathbf{\Sigma}(z)$; $\lambda_{\mathrm{high}} = \frac{1}{2}(\lambda_1+\lambda_2)$; $\bar{\lambda} = \frac{1}{8}\sum_m \lambda_m$; $e_m = \lambda_m/\sum_k \lambda_k$; $\hat{S}_z[k] = |Z[k]|^2/\sum_m |Z[m]|^2$; $W[k]$: DFT of interior residual $w = z - \hat{z}$ (4-sample margin); $P_u = \overline{|u|^2}$; $\epsilon = 10^{-8}$, $\epsilon' = 10^{-6}$; SD: standard deviation; IQR: interquartile range.}
\end{table}

\textbf{Autocorrelation Subspace Descriptors ($\Zsub$, 9-D):}
Clean modulated signals concentrate short-lag covariance energy in a few dominant eigenvalues $\lambda_1, \lambda_2$, whereas white noise lifts the spectral floor across all modes.
From biased autocorrelation lags $r_\ell(z) = \frac{1}{L}\sum_{n=\ell}^{L-1} z[n]z^*[n-\ell]$ ($\ell = 0,\dots,7$, $r_0 = 1$), we form the $8\times 8$ Hermitian positive-semidefinite Toeplitz matrix $[\mathbf{\Sigma}(z)]_{p,q} = r_{p-q}(z)$ ($p\ge q$), $[\mathbf{\Sigma}(z)]_{p,q} = r^*_{q-p}(z)$ ($p<q$) with eigenvalues $\lambda_1 \ge \cdots \ge \lambda_8 \ge 0$.
Drawing on eigenvalue spectrum sensing and effective rank~\cite{zeng2009eigenvalue,pauluzzi2000comparison,roy2007effective}, $\Zsub(z)\in\R^9$ pairs six spectral summaries with three time-domain summaries (standard deviation [SD], coherence, interquartile range [IQR]) of lag-1 phase differences $\Delta\phi[n]$ and envelopes $|z[n]|$~\cite{nandi1998algorithms} (Table~\ref{tab:descriptors}).
Because $(e^{j\phi}z[n])(e^{j\phi}z[n-\ell])^* = z[n]z^*[n-\ell]$, $|e^{j\phi}z[n]| = |z[n]|$, discrete Fourier transform (DFT) power spectra, and $\Delta\phi[n]$ cancel $e^{j\phi}$ identically, $\Zsub$ is \emph{strictly $U(1)$ phase-invariant} (Prior~P1).

\textbf{Blind-Spot Residual Descriptors ($\Zbs$, 6-D):}
To distinguish white noise from temporally correlated multipath or RF distortion without labels (Prior~P2), a 15.0K-parameter $J$-invariant 1D blind-spot network~\cite{batson2019noise2self,laine2019highquality}---split left/right-causal 3-layer dilated convolutions (32 channels, width 3, dilations $\{1,2,4\}$, 1-sample shift) fused by $1\times 1$ convolutions---predicts $\hat{z}[n]$ strictly from its 15-sample neighbors ($\partial \hat{z}[n]/\partial z[n] \equiv 0$).
Trained self-supervised via mean squared error on unlabeled source waveforms disjoint from $C_p$ plus $A$ under random $U(1)$ rotations, it absorbs predictable signal and correlated fading into $\hat{z}$ while forcing independent white noise into the residual $w = z - \hat{z}$.
Six energy, whiteness~\cite{ljung1978measure}, flatness, and correlation statistics of $(\hat{z}, w)$ form $\Zbs(z)\in\R^6$ (Table~\ref{tab:descriptors}), yielding $\Zsig = [\Zsub;\,\Zbs]\in\R^{15}$; $\hat{z}$ is purely diagnostic and never replaces $z$ at $f$'s input.

\textbf{Three Deployment Configurations:}
Under higher compute budgets, the rank model augments $\Zsig$ to 53-D using 15-D classical statistics~\cite{nandi1998algorithms,swami2000hierarchical} (envelope quantiles/moments, $|\overline{z^2}|$, cumulants $|C_{40}|, C_{42}$, $|r_\ell(z)|$ at $\ell\in\{1,2,4,8\}$, DFT entropy/peak) and 23-D \emph{active classifier probes} (12 extra passes, 13 total):
(i)~4-D $U(1)$ phase probes across $\phi\in\{\frac{\pi}{4},\frac{\pi}{2},\frac{3\pi}{4},\pi\}$ (mean Jensen--Shannon [JS] divergence, top-1 disagreement, log-probability SD, mutual information~\cite{gal2017bald}); and
(ii)~19-D AWGN probes~\cite{cohen2019smoothing} across $\eta\in\{0.1,0.316,1.0,3.162\}$ ($K_{\mathrm{mc}}=2$ draws/level: per-level entropy, JS, top-1 drop, pairwise total-variation [TV] distance, plus three summaries vs.\ $\log(1+\eta)$).
This yields three hardware-matched configurations (Section~\ref{sec:exp_edge}):
\textbf{PhyCal-Lite} (9-D $\Zsub$, $1\times$ pass, no denoiser; edge Internet of Things [IoT]),
\textbf{PhyCal-ZeroProbe} (15-D $\Zsig$, $1\times$ pass + 15.0K-parameter blind-spot denoiser; embedded software-defined radios [SDRs]), and
\textbf{PhyCal-Full} (53-D, $13\times$ passes; GPU base stations).

\subsection{Synthetic Channel Suite and Rank-Distribution Model (Module~2)}
\label{sec:module2}

To observe how frozen $f$ degrades without target labels (Prior~P2), Module~2 passes each $x_i^S \in C_p$ through a fixed $V = 8$-view unit-RMS \emph{synthetic channel suite}: one clean view ($v = 0$), five AWGN views ($v = 1,\dots,5$), and two causal two-tap Rician/CFO views ($v = 6,7$, mild/moderate; parameters in Section~\ref{sec:exp_setup}), forming $C_p^{(0)}$ and $C_p^{\mathrm{aug}}$.

Rather than recalibrating all $K$ class posteriors under severe shift, CP depends only on the ascending base-score rank $R(x,y)\in\{1,\dots,K\}$ of $s(x,y) = -\log p_f(y\mid x)$.
We model the conditional probability mass function (PMF) of the \emph{true-class rank} $R(x,Y)$---a smooth 1D ordinal variable governed primarily by signal quality.
To prevent in-sample optimism across views of the same frame $x_i^S$, $C_p$ is partitioned into five disjoint folds \emph{by base-frame identity} $k(i)\in\{1,\dots,5\}$~\cite{chernozhukov2018dml}.
Each fold fits a feature-wise linear modulation (FiLM)-conditioned \emph{rank model} ($\approx 13\text{K--}15\text{K}$ parameters: 64-unit sorted-score multilayer perceptron [MLP] modulated by a 32-unit physics MLP~\cite{perez2018film}) outputting rank PMF $q(r\mid x)\in\Delta^{K-1}$ and CDF $Q(c\mid x) = \sum_{r'=1}^c q(r'\mid x)$ by minimizing a \emph{tail-weighted Ranked Probability Score} (RPS) on $C_p^{\mathrm{aug}}$ plus an unlabeled Blackwell-inspired monotonicity penalty on $\mathcal{B}_A \subset A$:
\begin{equation}
\begin{split}
    \mathcal{L}(q) &= \E_{(x,y)\sim C_p^{\mathrm{aug}}}\!\Bigg[\frac{1}{K-1}\sum_{c=1}^{K-1}\omega_c(x) \\
    &\qquad \times \big(Q(c\mid x) - \Ind\{R(x,y)\le c\}\big)^2\Bigg] \\
    &\quad + \beta_{\mathrm{mono}}\,\E_{\mathcal{B}_A\subset A}\big[\ell_{\mathrm{BW}}(q; \mathcal{B}_A)\big],
\end{split}
\label{eq:head_loss}
\end{equation}
with $0.01$ rank cross-entropy stabilizer ($\beta_{\mathrm{mono}} = 0.5$).
Using $f$'s sorted cumulative probability $P_c(x) = \sum_{j=1}^c p_{(j)}(x)$, positive cutpoint weights focus on upper quantiles ($80\%\text{--}99\%$):
\begin{equation}
\begin{split}
    \tilde{\omega}_c(x) &= 1 + \gamma_{\mathrm{tail}}\exp\!\left(-\frac{(P_c(x) - \mu_{\mathrm{tail}})^2}{2\sigma_{\mathrm{tail}}^2}\right), \\
    \omega_c(x) &= \frac{(K-1)\,\tilde{\omega}_c(x)}{\sum_{c'=1}^{K-1}\tilde{\omega}_{c'}(x)},
\end{split}
\label{eq:twrps_weights}
\end{equation}
with $\gamma_{\mathrm{tail}} = 3.0$, $\mu_{\mathrm{tail}} = 0.90$, $\sigma_{\mathrm{tail}} = 0.12$.
Since $\omega_c(x) > 0$ depends solely on $p_f(\cdot\mid x)$ and not on $Y$, tail-weighted RPS is a \emph{strictly proper scoring rule} whose unique population minimizer is the exact conditional rank PMF~\cite{epstein1969scoring,gneiting2007strictly,gneiting2011weighted}.
Guided by Prior~P3, perturbing $x\in\mathcal{B}_A$ with AWGN ($\eta_{\mathrm{BW}}\in\{0.316, 1.0\}$) into $\tilde{x}$, $\ell_{\mathrm{BW}}$ penalizes batch-mean and per-frame violations of $Q(c\mid\tilde{x}) \le Q(c\mid x)$:
\begin{equation}
\begin{split}
    \ell_{\mathrm{BW}}(q;\mathcal{B}_A) &= \frac{1}{K-1}\sum_{c=1}^{K-1}\!\Bigg(\big[\bar{Q}_{\mathcal{B}}(c;\tilde{x}) - \bar{Q}_{\mathcal{B}}(c;x)\big]_+^2 \\
    &\quad + \frac{1}{4|\mathcal{B}_A|}\sum_{x\in\mathcal{B}_A}\big[Q(c\mid\tilde{x}) - Q(c\mid x) \\
    &\qquad\qquad\qquad - \delta_{\mathrm{BW}}\big]_+^2\Bigg),
\end{split}
\label{eq:bw_loss}
\end{equation}
where $[u]_+ = \max(0,u)$, $\bar{Q}_{\mathcal{B}}(c;\cdot)$ is the batch mean, and $\delta_{\mathrm{BW}} = 0.02$.
Because a frozen suboptimal $f$ can occasionally improve on an individual misclassified frame under noise, $\ell_{\mathrm{BW}}$ acts as a soft physics-inspired inductive bias rather than a hard pointwise law, motivating slack $\delta_{\mathrm{BW}} = 0.02$ and $1/4$ weight.

To remove mild conservative bias from tail weighting, a two-stage 5-fold OOF stacking step fits a view classifier $\hat{P}^{(-k)}(v\mid\zeta(x))$ and $2V$ per-view monotone temperature/tilt parameters $(a_v^{(-k)}, b_v^{(-k)})\in[-1.5,1.5]\times[-10,10]$ on the four complementary folds' Stage-1 OOF predictions by minimizing unweighted RPS~\cite{guo2017calibration}.
Setting $a^{(-k)}(x) = \sum_v \hat{P}^{(-k)}(v\mid\zeta(x))\,a_v^{(-k)}$ and $b^{(-k)}(x) = \sum_v \hat{P}^{(-k)}(v\mid\zeta(x))\,b_v^{(-k)}$ maps any input PMF $q_0$ via
\begin{equation}
\begin{split}
    \mathcal{T}^{(-k)}(q_0)(r\mid x) &\propto \exp\!\Big(e^{a^{(-k)}(x)}\log q_0(r\mid x) \\
    &\qquad\quad + b^{(-k)}(x)\frac{r-1}{K-1}\Big),
\end{split}
\label{eq:nested_recal}
\end{equation}
giving held-out source posteriors $\tilde{q}^{(-k)} = \mathcal{T}^{(-k)}(q^{(-k)})$ on $I_k$ and ensembled target posteriors $\tilde{q} = \frac{1}{5}\sum_{k=1}^5 \mathcal{T}^{(-k)}(\bar{q})$ on $U\cup T$, where $\bar{q} = \frac{1}{5}\sum_{j=1}^5 q^{(-j)}$.
This defines rank-indexed class probabilities $\tilde{p}(y\mid x) \triangleq \tilde{q}(R(x,y)\mid x)$ and predicted score CDF $G_{\tilde{q}}(x,t) \triangleq \sum_{y\in\calY} \tilde{p}(y\mid x)\Ind\{\tilde{s}(x,y)\le t\}$; all three nonconformity scores below preserve $f$'s class ordering.

\subsection{Channel-Mixture Transport and Doubly Robust Calibration (Modules~3 \& 4)}
\label{sec:module34}

\textbf{Degradation-Gated View-Simplex Expectation--Maximization (Module~3):}
Rather than reweighting individual frames via high-dimensional neural density ratios as in WCP~\cite{tibshirani2019wcp} (which collapses $\ESS$ under SNR drops), Module~3 transports at the \emph{channel-mixture level} in classifier-free descriptor space $\zeta(x) \triangleq \Zsig(x)$ ($\Zsub(x)$ in PhyCal-Lite), writing $\zeta_{v,i}^S = \zeta(x_{v,i}^S)$ and $\zeta_a^A = \zeta(x_a^A)$.
Because the synthetic suite degrades strictly forward (Prior~P2), we compute the standardized median SNR shift from clean $C_p^{(0)}$ to $A$ over log blind-SNR coordinate(s) $\mathcal{S}$ of $\zeta$ (Toeplitz proxy in Lite, averaged with blind-spot log SNR in ZeroProbe/Full):
\begin{equation}
    \Delta_{\mathrm{snr}} = \frac{1}{|\mathcal{S}|}\sum_{j\in\mathcal{S}} \frac{\mathrm{median}_{a\in A}(\zeta_{a,j}^A) - \mathrm{median}_{i\in C_p^{(0)}}(\zeta_{0,i,j}^S)}{\mathrm{SD}_{i\in C_p^{(0)}}(\zeta_{0,i,j}^S)}.
    \label{eq:delta_snr}
\end{equation}
If $\Delta_{\mathrm{snr}} \ge -0.25$~SD (stationary or cleaner links), a one-way \emph{degradation gate} selects $\hat{\pi} = e_0 = (1,0,\dots,0)^\top$ so noise views never contaminate stationary calibration.
Otherwise ($\Delta_{\mathrm{snr}} < -0.25$~SD), an $L_2$-regularized MLP view discriminator $\hat{P}(v\mid\zeta)$ across equal-prior source views drives maximum-likelihood \emph{View-Simplex expectation--maximization (EM)}~\cite{saerens2002adjusting,alexandari2020mlls} on $A$ from $\pi_v^{(0)} = 1/V$:
\begin{equation}
    \gamma_{a,v}^{(\nu)} = \frac{\pi_v^{(\nu)} \hat{P}(v\mid \zeta_a^A)}{\sum_{v'=0}^{V-1}\pi_{v'}^{(\nu)} \hat{P}(v'\mid \zeta_a^A)},\quad
    \pi_v^{(\nu+1)} = \frac{1}{N_A}\sum_{a=1}^{N_A} \gamma_{a,v}^{(\nu)},
    \label{eq:view_em}
\end{equation}
yielding $\hat{\pi}\in\Delta^{V-1}$.
Distributing each view's mass uniformly via $w_{v,i}^* = \hat{\pi}_v / N_{\mathrm{base}}$ matches the channel mixture while preserving maximal effective sample sizes $\ESS_{\mathrm{base}} = \ESS_v = N_{\mathrm{base}}$ (Proposition~\ref{prop:em_ess}).

\textbf{Doubly Robust CDF Calibration (Module~4):}
With $\tilde{q}$ and $w_{v,i}^*$ fixed from $(C_p, A)$, Module~4 estimates $F_T(t)$ via a channel-matched doubly robust (DR) estimator~\cite{robins1994estimation,yang2024doubly,angelopoulos2023ppi} that sums the initial target CDF estimate on fresh $U$ and the channel-matched OOF source debiasing residual:
\begin{equation}
\begin{split}
    \widetilde{F}_T^{\mathrm{raw}}(t) &= \frac{1}{N_U}\sum_{x\in U} G_{\tilde{q}}(x,t) + \sum_{v=0}^{V-1}\sum_{i=1}^{N_{\mathrm{base}}} w_{v,i}^* \\
    &\quad \times \Big(\Ind\{\tilde{s}(x_{v,i}^S, y_i^S)\le t\} - G_{\tilde{q}}^{(-k(i))}(x_{v,i}^S, t)\Big).
\end{split}
\label{eq:pbdr_master}
\end{equation}
Clipping $\widetilde{F}_T^{\mathrm{raw}}(t)$ to $[0,1]$ and applying isotonic projection yields a valid CDF $\widetilde{F}_T(t)$ with finite-sample-adjusted calibration threshold
\begin{equation}
    \hat{\tau}_\alpha = \inf\!\left\{ t\in\R : \widetilde{F}_T(t) \ge \gamma_n \right\},
    \label{eq:threshold_inv}
\end{equation}
where $\gamma_n = \min\!\big(1,\, (1-\alpha)(1 + N_U^{-1})\big)$.

\textbf{Three Scoring Modes:}
Evaluating~\eqref{eq:pbdr_master}--\eqref{eq:threshold_inv} under three scores serves complementary goals (denoted PhyCal-\{Full, ZeroProbe, Lite\}-\{Marg, SNR, Piv\}):
\begin{enumerate}
    \item \textbf{PhyCal-Marg} (\emph{smallest sets}): $\tilde{s}_{\mathrm{Marg}}(x,y) = s(x,y) = -\log p_f(y\mid x)$~\cite{sadinle2019least}, a global cutoff minimizing $\ASS$ under balanced traffic.
    \item \textbf{PhyCal-SNR} (\emph{SNR uniformity, deterministic order}): $\tilde{s}_{\mathrm{SNR}}(x,y) = \log(1 + s(x,y)) - \hat{b}_{1-\alpha}(\zeta(x))$, where $\hat{b}_{1-\alpha}$ is a 5-fold OOF non-crossing pinball MLP~\cite{koenker1978regression,romano2019cqr} predicting the $(1-\alpha)$-quantile of $\log(1+s(X,Y))$ solely from $\zeta(x)$, removing SNR difficulty while preserving $f$'s class order.
    \item \textbf{PhyCal-Piv} (\emph{SNR uniformity \& hard-class protection}): inspired by DCP~\cite{chernozhukov2021dcp}, $\tilde{s}_{\mathrm{Piv}}(x,y) = \sum_{r'=1}^{R(x,y)-1} \tilde{q}(r'\mid x) + u_x\,\tilde{q}(R(x,y)\mid x) \in [0,1]$ with $u_x \sim \mathrm{Unif}(0,1)$ ($\tilde{q}^{(-k(i))}$ on fold $k(i)$), a randomized probability integral transform (PIT) that is conditionally $\mathrm{Unif}(0,1)$ given $X$ when $\tilde{q}$ is well-specified, equalizing per-SNR coverage and empirically lifting hard-class coverage.
\end{enumerate}

\textbf{Label-Free Deployment Selection:}
Configuration and mode are chosen offline without target labels: use \textbf{PhyCal-Lite} on edge CPUs or unadapted \emph{-Src} models, \textbf{PhyCal-ZeroProbe} under correlated OTA fading ($+8\%\text{--}9\%$ MACs), and \textbf{PhyCal-Full} on GPUs wrapping adapted models; use \textbf{PhyCal-Piv} by default for SNR uniformity and hard-class protection, \textbf{PhyCal-SNR} for deterministic ordering or equal-SNR fading ($\Delta_{\mathrm{snr}}\approx 0$), and \textbf{PhyCal-Marg} for minimal $\ASS$ under balanced traffic (Algorithm~\ref{alg:phycal}).

\begin{algorithm}[!t]
\caption{PhyCal Unlabeled Calibration and Inference}
\label{alg:phycal}
\footnotesize
\begin{algorithmic}[1]
\Require Frozen $f$; labeled $C_p$; unlabeled $A, U$; stream $T$; $\alpha$; config $\in \{\text{Lite}, \text{ZeroProbe}, \text{Full}\}$; mode $\in \{\text{Marg}, \text{SNR}, \text{Piv}\}$.
\Ensure Online calibrated prediction sets $\{\calC_\alpha(x)\}_{x\in T}$.
\Statex \textbf{Phase 1: Offline Calibration on $(C_p, A)$ (Zero Target Labels)}
\State Generate $V = 8$ unit-RMS views $C_p^{\mathrm{aug}}$ from $C_p$ (Sec.~\ref{sec:module2}).
\State Extract $\zeta(\cdot)$ via M1 (Table~\ref{tab:descriptors}; fit 15.0K denoiser if ZeroProbe/Full; add active probes and classical statistics if Full).
\State Split $C_p$ into 5 base-frame folds $\{I_k\}$; fit FiLM rank models~\eqref{eq:head_loss}--\eqref{eq:bw_loss} and OOF RPS stacking~\eqref{eq:nested_recal} for $\{\tilde{q}^{(-k)}\}, \tilde{q}$.
\State Compute $\Delta_{\mathrm{snr}}$ via~\eqref{eq:delta_snr}:
\If{$\Delta_{\mathrm{snr}} \ge -0.25$~SD}
    \State Degradation gate: $\hat{\pi} \gets e_0$.
\Else
    \State Fit $\hat{P}(v\mid\zeta)$ on $C_p^{\mathrm{aug}}$; run View-Simplex EM~\eqref{eq:view_em} on $A$ for $\hat{\pi}$.
\EndIf
\State Set uniform weights $w_{v,i}^* \gets \hat{\pi}_v / N_{\mathrm{base}}$ ($\ESS_{\mathrm{base}} = N_{\mathrm{base}}$).
\State (If mode = SNR) Fit 5-fold OOF quantile MLP $\hat{b}_{1-\alpha}(\zeta(\cdot))$ on $C_p^{\mathrm{aug}}$.
\Statex \textbf{Phase 2: Threshold Estimation on Independent Window $U$}
\State Compute $\tilde{s}$, $G_{\tilde{q}}$, and $\widetilde{F}_T^{\mathrm{raw}}(t)$ via~\eqref{eq:pbdr_master}; clip, project isotonically to $\widetilde{F}_T$, and invert~\eqref{eq:threshold_inv} for $\hat{\tau}_\alpha$.
\Statex \textbf{Phase 3: Real-Time Online Inference on Stream $T$}
\For{each incoming frame $x \in T$}
    \State Compute $p_f(\cdot\mid x)$, $\zeta(x)$, and $\tilde{s}(x,y)$ ($y\in\calY$); output $\calC_\alpha(x) = \{y : \tilde{s}(x,y) \le \hat{\tau}_\alpha\}$.
\EndFor
\end{algorithmic}
\end{algorithm}

\subsection{Theoretical Guarantees and Identifiability Limits}
\label{sec:theory}

Two propositions, one remark, and one corollary establish PhyCal's optimization optimality, doubly robust error decomposition, cross-fitted score extension, assumption-dependent approximate coverage control, and physical identifiability boundaries (complete proofs in Appendices~\ref{app:prop1}--\ref{app:cor1}).

\begin{proposition}[EM Global Concavity and Maximal Source ESS]
\label{prop:em_ess}
Let the $V$ source views induce equal-prior densities $\{p_{S,v}(\zeta)\}_{v=0}^{V-1}$ with mixture average $\bar{p}_S(\zeta) = \frac{1}{V}\sum_{v=0}^{V-1} p_{S,v}(\zeta)$, and suppose the view discriminator satisfies the Bayes-consistent ratio $\hat{P}(v\mid \zeta) = p_{S,v}(\zeta)/(V\bar{p}_S(\zeta))$. Then:
\begin{enumerate}
    \item The EM updates~\eqref{eq:view_em} monotonically ascend the concave log-likelihood $\mathcal{J}(\pi) = \frac{1}{N_A}\sum_{a=1}^{N_A}\log\!\big(\sum_{v=0}^{V-1} \pi_v \hat{P}(v\mid \zeta_a^A)\big)$ to a global maximizer $\hat{\pi}\in\Delta^{V-1}$.
    \item For any non-negative $\{w_{v,i}\}$ with $\sum_{i=1}^{N_{\mathrm{base}}} w_{v,i} = \hat{\pi}_v$, base-frame marginals $W_i = \sum_v w_{v,i}$, and effective sample sizes
    \[
    \begin{aligned}
        \ESS_{\mathrm{base}}(w) &= \frac{\big(\sum_i W_i\big)^2}{\sum_i W_i^2}, \\
        \ESS_v(w) &= \frac{\big(\sum_i w_{v,i}\big)^2}{\sum_i w_{v,i}^2}\quad (\hat{\pi}_v > 0),
    \end{aligned}
    \]
    uniform weights $w_{v,i}^* = \hat{\pi}_v / N_{\mathrm{base}}$ uniquely attain both maxima simultaneously: $\ESS_{\mathrm{base}}(w^*) = \ESS_v(w^*) = N_{\mathrm{base}}$.
\end{enumerate}
\end{proposition}

\begin{IEEEproof}
See Appendix~\ref{app:prop1}.
\end{IEEEproof}

Three finite-sample caveats apply: $\mathcal{J}(\pi)$ remains concave for any positive neural $\hat{P}(v\mid\zeta)$, whereas exact target-mixture consistency requires Bayes-calibrated $\hat{P}$~\cite{alexandari2020mlls}; uniform weights $w_{v,i}^*$ maximize $\ESS$ as a variance--bias trade-off absorbing any within-view mismatch into Term~(III) of Proposition~\ref{prop:dr_decomp}; and matching the marginal $p_T(\zeta)$ on $A$ targets channel-induced quality shift rather than guaranteeing full joint equality $P_T(X,Y)=\sum_v \hat{\pi}_v P_{S,v}(X,Y)$.

\begin{proposition}[Doubly Robust CDF Error Decomposition and Boundary]
\label{prop:dr_decomp}
For a fixed frozen classifier $f$, fixed score function $\tilde{s}$ (e.g., PhyCal-Marg), 5-fold balanced OOF estimators $\{\tilde{q}^{(-k)}\}_{k=1}^5$ fitted strictly on $(C_p^{(-k)}, A)$ with average $\tilde{q} = \frac{1}{5}\sum_{k=1}^5 \tilde{q}^{(-k)}$, and mixture $\hat{\pi}\in\Delta^{V-1}$ fitted on $(C_p, A)$, define for any distribution $D$ the target CDF $F_D(t) = \Prob_D\{\tilde{s}(X,Y)\le t\}$, predicted CDF $\mu_D(t) = \E_D[G_{\tilde{q}}(X,t) \mid C_p, A]$, and residual $\delta_D(t) = F_D(t) - \mu_D(t)$, with source-mixture residual $\delta_{S,\hat{\pi}}(t) = \sum_{v=0}^{V-1}\hat{\pi}_v \delta_{S,v}(t)$.
Denoting the two sums in~\eqref{eq:pbdr_master} by $\widehat{\mu}_U(t)$ and $\widehat{\delta}_{S,\hat{\pi}}(t) = \sum_{v=0}^{V-1}\hat{\pi}_v \widehat{\delta}_{S,v}(t)$, every $t\in\R$ satisfies
\begin{equation}
\begin{split}
    \widetilde{F}_T^{\mathrm{raw}}(t) - F_T(t) &= \underbrace{\big(\widehat{\mu}_U(t) - \mu_T(t)\big)}_{\text{(I) Target } U \text{ sampling } O_p(N_U^{-1/2})} \\
    &\quad + \underbrace{\big(\widehat{\delta}_{S,\hat{\pi}}(t) - \delta_{S,\hat{\pi}}(t)\big)}_{\text{(II) Source OOF residual } O_p(N_{\mathrm{base}}^{-1/2})} \\
    &\quad + \underbrace{\big(\delta_{S,\hat{\pi}}(t) - \delta_T(t)\big)}_{\text{(III) Residual transport bias}}.
\end{split}
\label{eq:dr_decomp}
\end{equation}
In particular:
\begin{itemize}
    \item[(a)] \emph{Double robustness}: Term~(III) vanishes if \emph{either} $\tilde{p}(y\mid x)$ is well-specified on $P_T$ and active views $\hat{\pi}_v > 0$ ($\delta_T = \delta_{S,\hat{\pi}} = 0$), \emph{or} $P_T = \sum_{v=0}^{V-1}\pi_v^* P_{S,v}$ lies in the source-view simplex with $\hat{\pi}\to\pi^*$ even under misspecified $\tilde{q}$.
    \item[(b)] \emph{Forward-degradation boundary}: When $P_T$ lies outside the convex hull of $\{P_{S,v}\}_{v=0}^{V-1}$ (e.g., reverse near-field OTA drift with $\Delta_{\mathrm{snr}}>0$, or long-delay multipath exceeding the finite-lag observation window) and $\tilde{q}$ is misspecified, no $\hat{\pi}\in\Delta^{V-1}$ guarantees $\delta_{S,\hat{\pi}}(t) = \delta_T(t)$ without target labels unless $f$'s representation absorbs the shift.
\end{itemize}
\end{proposition}

\begin{IEEEproof}[Sketch]
Adding and subtracting $\delta_{S,\hat{\pi}}$ yields~\eqref{eq:dr_decomp}.
Conditioned on $(C_p,A)$ and each complementary fold $(C_p^{(-k)},A)$, monotonicity of $G_{\tilde{q}}$ and threshold indicators (VC-subgraph dimension~1) lifts finite-sample bounds uniformly over $t\in\R$, while 5-fold linearity and a finite-simplex union bound give uniform $O_p(N_U^{-1/2})$ and $O_p(N_{\mathrm{base}}^{-1/2})$ rates for Terms~(I)--(II) under strict fold-wise independence (approximated in implementation by low-DOF $2V$-parameter Stage-2 stacking~\eqref{eq:nested_recal}; Appendix~\ref{app:prop2}); double robustness and the out-of-hull boundary follow from linearity over $\{P_{S,v}\}$.
\end{IEEEproof}

\begin{remark}[Extension to Cross-Fitted Adaptive Scores]
\label{rem:adaptive_scores}
For PhyCal-SNR and PhyCal-Piv, target $U\cup T$ uses the 5-fold ensemble score $\tilde{s}$, whereas held-out fold $I_k$ evaluates the independent OOF score $\tilde{s}^{(-k)}$ fitted on $(C_p^{(-k)}, A)$.
Uniform $O_p(N_U^{-1/2})$ and $O_p(N_{\mathrm{base}}^{-1/2})$ concentration still holds around the OOF average $\bar{\delta}_{S,\hat{\pi}}^{\mathrm{OOF}}(t)$, adding one explicit cross-fold stability remainder $\Delta_{\mathrm{ens}}(t) = \bar{\delta}_{S,\hat{\pi}}^{\mathrm{OOF}}(t) - \delta_{S,\hat{\pi}}(t)$ to~\eqref{eq:dr_decomp}; under PhyCal-Piv, randomized PIT cancels $\mu_{S,v}^{(-k)}(t) = \mu_{S,v}(t) = t$ identically on $[0,1]$ (Appendix~\ref{app:prop2}).
\end{remark}

\begin{corollary}[Isotonic Non-Expansiveness and Approximate Coverage Control]
\label{cor:coverage_bound}
Let $\varepsilon_n = \sup_{t\in\R} |\widetilde{F}_T^{\mathrm{raw}}(t) - F_T(t)|$ denote the uniform CDF error prior to post-processing, and let $\gamma_n = \min(1, (1-\alpha)(1 + N_U^{-1}))$ be the finite-sample-adjusted target level in~\eqref{eq:threshold_inv}.
Then the clipped, isotonically projected CDF satisfies $\|\widetilde{F}_T - F_T\|_\infty \le \varepsilon_n$, and the operating threshold $\hat{\tau}_\alpha$ guarantees conditional-on-calibration target coverage
\begin{equation}
    \Prob_{P_T}\{Y \in \calC_\alpha(X) \mid C_p, A, U\} = F_T(\hat{\tau}_\alpha) \ge \gamma_n - \varepsilon_n.
    \label{eq:cov_lower}
\end{equation}
Moreover, whenever $F_T$ is continuous (e.g., under randomized PhyCal-Piv), two-sided error satisfies $|F_T(\hat{\tau}_\alpha) - (1-\alpha)| \le \varepsilon_n + (1-\alpha)/N_U$.
\end{corollary}

\begin{IEEEproof}
See Appendix~\ref{app:cor1}.
\end{IEEEproof}

Operationally, Proposition~\ref{prop:dr_decomp}, Remark~\ref{rem:adaptive_scores}, and Corollary~\ref{cor:coverage_bound} provide \emph{assumption-dependent approximate coverage control} rather than distribution-free validity under arbitrary shift (unachievable without target labels by Proposition~\ref{prop:dr_decomp}(b)): under i.i.d.\ sampling within $C_p$ and $U$, $\varepsilon_n \le O_p(N_U^{-1/2} + N_{\mathrm{base}}^{-1/2}) + \|\delta_{S,\hat{\pi}} - \delta_T\|_\infty + \|\Delta_{\mathrm{ens}}\|_\infty$, where the transport bias $\|\delta_{S,\hat{\pi}} - \delta_T\|_\infty$ vanishes if $\tilde{q}$ generalizes accurately or $P_T$ lies in the eight-view convex hull with $\hat{\pi}\to\pi^*$, directly predicting the out-of-hull regimes in Section~\ref{sec:experiments}.

% =============================================================================
\section{Experimental Evaluation}
\label{sec:experiments}
% =============================================================================

% -----------------------------------------------------------------------------
\subsection{Experimental Setup}
\label{sec:exp_setup}

\begin{table*}[!t]
\centering
\caption{Within-Dataset Wireless Channel-Shift Benchmarks}
\label{tab:benchmarks}
\footnotesize
\renewcommand{\arraystretch}{1.0}
\setlength{\tabcolsep}{3pt}
\begin{tabular*}{\textwidth}{@{\extracolsep{\fill}}llcll@{}}
\toprule
\textbf{Benchmark} & \textbf{RF Platform} & \textbf{$K / L$} & \textbf{Source $\leftrightarrow$ Target} & \textbf{Physical Shift Nature} \\
\midrule
\textbf{RML22}~\cite{sathyanarayanan2023rml22} & Synthetic RF channel & 11 / 128 & High ($10\text{--}18$~dB) $\leftrightarrow$ Low ($-6\text{--}0$~dB) & Controlled steep SNR shift \\
\textbf{MIGOU-MOD}~\cite{utrilla2020migou} & Real USRP / MIGOU OTA & 11 / 128 & Near ($1$~m) $\leftrightarrow$ Far ($6$~m) & Coupled distance, RF, and SNR$^\dagger$ \\
\textbf{Multipath OTA}~\cite{belousov2026multipath} & Real 2.4~GHz HackRF One & 7$^\ddagger$ / \textbf{1{,}024} & LoS $\leftrightarrow$ 3-tap MP ($[0,100,200]\,\mu$s) & Equal-SNR ($20\text{--}30$~dB) long-delay ISI \\
\bottomrule
\end{tabular*}
\vspace{2pt}
\parbox{\textwidth}{\scriptsize $K$ and $L$ denote modulation classes and native I/Q frame length; H/L, MP, and MIGOU abbreviate High/Low, multipath, and MIGOU-MOD in shift labels. $^\dagger$Distance change ($1$~m vs.\ $6$~m) couples propagation, RF hardware, and SNR shifts. $^\ddagger$Indoor 3~m OTA link with OFDM, 16-QAM, GMSK, BPSK, QPSK, NBFM, and WBFM across official disjoint train/validation/test splits (132{,}448 frames).}
\end{table*}

\subsubsection{Benchmarks, Splits, and Metrics}
We evaluate PhyCal on three public AMC benchmarks spanning synthetic SNR drops, real Universal Software Radio Peripheral (USRP) IoT OTA links, and real 2.4~GHz HackRF One long-delay intersymbol interference (ISI) multipath (Table~\ref{tab:benchmarks};~\cite{sathyanarayanan2023rml22,utrilla2020migou,belousov2026multipath}).
Matching Section~\ref{sec:prob_setup}, each domain uses disjoint unit-RMS splits ($N_{\mathrm{base}} = N_A = N_U = 1{,}024$, $N_T = 3{,}072$) across all four pairings (two cross-domain, two in-domain) and 3 seeds $\times$ 10 batches/seed (92{,}160 frame evaluations per scenario across 10 target test batches reused across seeds).
One fixed 8-view synthetic channel suite---clean, five AWGN views ($\eta\in\{0.1, 0.316, 1.0, 2.0, 4.0\}$), and two causal two-tap Rician/CFO views (mild: tap $[0.10,0.35]$, $|\mathrm{CFO}|\le 0.008$~cycles/sample, $\eta = 0.03$; moderate: $[0.20,0.50]$, $|\mathrm{CFO}|\le 0.015$, $\eta = 0.25$)---and all hyperparameters are shared across benchmarks without per-dataset tuning.

Shifts are denoted $\text{Source}\to\text{Target}$, with \emph{Cross} averaging both cross-domain shifts of a benchmark; \emph{4C}/\emph{8S} denote the four cross-domain shifts / all eight scenarios on 11-class $L = 128$ benchmarks (RML22 + MIGOU-MOD), and \emph{6 In-Dom.}\ spans all six stationary controls across all three benchmarks.
Over $1-\alpha \in \{75\%, 80\%, 85\%, 90\%, 95\%, 99\%\}$ (jointly predicted by PhyCal-SNR), we report on $T$ empirical coverage $\widehat{\mathrm{Cov}} = \frac{1}{N_T}\sum_{n=1}^{N_T}\Ind\{y_n^T \in \calC_\alpha(x_n^T)\}$, batch-wise mean absolute error (MAE), $\MAE = \E[|\widehat{\mathrm{Cov}} - (1-\alpha)|]$, in percentage points (pp), $\ASS$, saving~\eqref{eq:demod_saving}, unobserved per-SNR spread $\max_g \widehat{\mathrm{Cov}}_g - \min_g \widehat{\mathrm{Cov}}_g$, and worst-class coverage $\min_y \widehat{\mathrm{Cov}}(Y = y)$, observing zero target labels or target SNR metadata during calibration ($T$ labels are used only offline).

\subsubsection{Classifier Matrix and Baselines}
To verify classifier agnosticism, we freeze a $3\times 2$ matrix of three AMC architectures---1D ResNet-18~\cite{he2016resnet}, hybrid MCLDNN~\cite{xu2020mcldnn}, and complex wavelet--bidirectional gated recurrent unit CSPMNet~\cite{zhang2026cspmnet} (241K parameters)---under \emph{source-only} (\emph{-Src}) and \emph{unsupervisedly adapted} (\emph{-Adapt}) regimes.
Each \emph{-Adapt} model is warm-started from its source checkpoint, adapted on unlabeled $A$ for 3 epochs via CDAN~\cite{long2018cdan}, and paired with a source-trained two-layer MLP recalibration head (waveform-quality branch isolated from domain gradients), then frozen prior to calibration.
On every model we benchmark 12 unlabeled baselines grouped by access (Table~\ref{tab:baselines_main}):
\emph{source-only ($C_p$)}---Source-LAC, Source-APS, Source-RAPS;
\emph{target-only ($U$, plus an extra source-trained same-architecture teacher for SFCP-Ind)}---Plug-in Softmax ($p_f$ on $U$, $w_{v,i}^* = 0$ in~\eqref{eq:pbdr_master}) and SFCP-Ind~\cite{angelman2025sfcp}; and
\emph{source + unlabeled target ($C_p, A/U$)}---WCP, Feature-DR~\cite{qiu2023adaptive,yang2024doubly} (WCP weights + raw softmax), ECP, QTC, OT-CP, and I/Q-Mixup temperature scaling~\cite{hu2024pseudocal} on $A$ via $-\log p_f$ CDF thresholding (PseudoCal-Soft) or cumulative scaled sets (PseudoCal-TopP).
PhyCal-Lite uses $(f, C_p, A, U)$ with synthetic source views; PhyCal-ZeroProbe/Full add the 15.0K-parameter self-supervised blind-spot denoiser on 2{,}048 disjoint unlabeled source frames plus $A$.
Three target-supervised references (Target-LAC*/APS*/RAPS*, $N_U = 1{,}024$) serve solely as labeled target reference benchmarks.

\textbf{Implementation Details:}
Rank models (64/32-unit FiLM MLPs) use AdamW ($\mathrm{lr} = 2\times 10^{-3}$, weight decay $10^{-4}$, batch 256, 25--80 epochs, 8-epoch warmup, patience 10 on source validation, $\beta_{\mathrm{mono}} = 0.5$, $\delta_{\mathrm{BW}} = 0.02$, $0.01$ cross-entropy weight); Stage-2 OOF RPS uses L-BFGS-B ($\le 200$ iterations, $L_2 = 10^{-5}$).
The 15.0K-parameter blind-spot denoiser uses the same AdamW settings (30 epochs, batch 128; 20 epochs, batch 64 on Multipath OTA); view MLPs (32 SiLU, 120 epochs, $\mathrm{lr} = 10^{-2}$, decay $0.05$, cosine) drive EM ($\le 200$ iterations, tolerance $10^{-8}$); pinball MLPs (32--16 SiLU) train for 80 epochs (batch 512, $\mathrm{lr} = 5\times 10^{-3}$, decay $10^{-3}$).
Baseline defaults: WCP/Feature-DR logistic regression ($C = 0.5$, weights $[10^{-3},100]$), RAPS ($k_{\mathrm{reg}} = 2,\lambda = 0.05$), PseudoCal Mixup $0.65/0.35$ (temperature in $[0.05, 20]$), OT-CP bandwidth $0.1$.
All runs use PyTorch~2.12.1 on an RTX 2080 Ti GPU and Intel Core i7-7800X (3.50~GHz) CPU with deterministic replay.

% -----------------------------------------------------------------------------
\subsection{Calibration Accuracy Across Baselines and Classifier Architectures}
\label{sec:exp_core}

\begin{table*}[!t]
\centering
\caption{Comparison Against 12 Unlabeled Baselines and 3 Supervised References on Frozen ResNet-18-Adapt (RML22 + MIGOU-MOD)}
\label{tab:baselines_main}
\renewcommand{\arraystretch}{0.93}
\setlength{\tabcolsep}{4.2pt}
\resizebox{\textwidth}{!}{%
\begin{tabular}{@{}llccccccccc@{}}
\toprule
\multirow{2}{*}{\textbf{Method}} & \multirow{2}{*}{\textbf{Category / Source}} & \multicolumn{6}{c}{\textbf{Nominal Coverage $1-\alpha = 90\%$}} & \multicolumn{3}{c}{\textbf{High-Rel.\ $1-\alpha = 99\%$}} \\
\cmidrule(lr){3-8} \cmidrule(lr){9-11}
 & & \textbf{4C-MAE$\downarrow$} & \textbf{8S-MAE$\downarrow$} & \textbf{4C-ASS} & \textbf{4C-WC$\uparrow$} & \textbf{H$\to$L Cov.} & \textbf{H$\to$L Spr.$\downarrow$} & \textbf{4C-MAE$\downarrow$} & \textbf{H$\to$L ASS} & \textbf{Sav.(\%)$\uparrow$} \\
\midrule
Source-LAC & Source CP~\cite{sadinle2019least} & 10.94 & 6.05 & 1.38 & 51.7\% & 56.8\% & 28.4 & 3.67 & 2.43 & 77.9$^\dagger$ \\
Source-APS & Adaptive Source~\cite{romano2020aps} & 2.80 & 1.76 & 1.88 & 70.5\% & 82.4\% & 12.8 & 1.57 & 3.81 & 65.3$^\dagger$ \\
Source-RAPS & Reg.\ Source~\cite{angelopoulos2021raps} & 4.97 & 2.88 & 1.77 & 69.0\% & 78.0\% & 16.5 & 3.95 & 2.59 & 76.5$^\dagger$ \\
Plug-in Softmax & Plug-in CDF ($p_f$, $w^*{=}0$) & 5.66 & 3.96 & 1.49 & 55.6\% & 81.1\% & 21.8 & 1.62 & 3.80 & 65.4$^\dagger$ \\
WCP & Feat.\ Density Ratio~\cite{tibshirani2019wcp} & 7.23 & 4.58 & 1.39 & 55.8\% & 66.0\% & 23.7 & 1.71 & 6.72 & 38.9$^\dagger$ \\
Feature-DR & Feature-Space DR~\cite{qiu2023adaptive,yang2024doubly} & 5.52 & 3.66 & 1.45 & 58.4\% & 75.0\% & 25.0 & 3.18 & 2.60 & 76.4$^\dagger$ \\
ECP & Entropy-Scaled~\cite{kasa2025ecp} & 7.90 & 5.00 & 1.54 & 59.0\% & 70.2\% & 27.4 & 2.36 & 3.05 & 72.3$^\dagger$ \\
QTC & Test Quantile~\cite{yilmaz2022qtc} & 2.79 & 2.00 & 2.28 & 68.2\% & 95.0\% & 8.5 & 0.63 & 11.00 & 0.0 (full) \\
OT-CP & Optimal Transport~\cite{correia2025otcp} & 40.51 & 37.64 & 0.59 & 3.0\% & 54.6\% & 27.8 & 27.31 & 2.17 & 80.3$^\dagger$ \\
PseudoCal-Soft & Mixup Temp.\ CDF~\cite{hu2024pseudocal} & 8.85 & 9.33 & 4.84 & 96.1\%$^\ddagger$ & 98.2\% & 3.9$^\ddagger$ & 0.97 & 9.13 & 17.0 \\
PseudoCal-TopP & Mixup Cumul.\ Prob.~\cite{hu2024pseudocal} & 9.40 & 9.62 & 5.42 & 97.6\%$^\ddagger$ & 98.3\% & 3.4$^\ddagger$ & 0.98 & 9.77 & 11.2 \\
SFCP-Ind & Pseudo-Label CP~\cite{angelman2025sfcp} & 4.71 & 3.86 & 2.76 & 70.8\% & 99.0\% & 2.3$^\ddagger$ & 0.70 & 8.93 & 18.8 \\
\midrule
\textbf{PhyCal-Lite-SNR} & \textbf{Ours (0.3\% MACs)} & \underline{0.97} & \underline{1.16} & 2.04 & 68.2\% & 89.8\% & \underline{5.4} & \underline{0.53} & \underline{5.09} & \underline{53.8} \\
\textbf{PhyCal-Full-Marg} & \textbf{Ours (Global $-\log p_f$)} & 1.55 & 1.42 & \textbf{1.86} & \underline{72.0\%} & 89.8\% & 15.4 & 0.89 & 6.09 & 44.6 \\
\textbf{PhyCal-Full-SNR} & \textbf{Ours (Signal Quantile)} & 1.41 & 1.29 & 2.15 & 70.0\% & 89.8\% & 5.5 & 0.59 & \textbf{5.07} & \textbf{53.9} \\
\textbf{PhyCal-Full-Piv} & \textbf{Ours (CDF Pivotal)} & \textbf{0.84} & \textbf{0.90} & 2.33 & \textbf{76.0\%} & \textbf{89.9\%} & \textbf{2.6} & \textbf{0.37} & 5.31 & 51.8 \\
\midrule
Target-LAC* ($N_U{=}1{,}024$) & Labeled Target Ref. & 0.93 & 0.91 & 1.82 & 69.8\% & 90.1\% & 15.1 & 0.29 & 6.40 & 41.8 \\
Target-APS* ($N_U{=}1{,}024$) & Labeled Target Ref. & 0.80 & 0.82 & 2.07 & 74.4\% & 90.1\% & 10.9 & 0.26 & 6.19 & 43.8 \\
Target-RAPS* ($N_U{=}1{,}024$) & Labeled Target Ref. & 0.85 & 0.82 & 2.09 & 73.3\% & 90.2\% & 14.1 & 0.30 & 8.89 & 19.2 \\
\bottomrule
\end{tabular}}
\vspace{1pt}
\parbox{\textwidth}{\scriptsize Best calibrated unlabeled result in \textbf{bold}, second best \underline{underlined} (excluding miscalibrated methods marked $\dagger$/$\ddagger$, as well as 90\% under-covering rows on 4C-ASS). 4C / 8S: four cross-domain shifts / all eight scenarios of RML22 + MIGOU-MOD; WC: worst-class coverage; H$\to$L Cov., H$\to$L Spr., and Sav.\ denote RML22 H$\to$L empirical coverage, SNR-conditional coverage spread (pp), and candidate-hypothesis reduction proxy~\eqref{eq:demod_saving}; MAE in pp. $^\dagger$Severe 99\% under-coverage ($76.8\%\text{--}94.5\%$); saving invalid. $^\ddagger$Distorted by $98\%\text{--}99\%$ over-coverage. *Uses $N_U = 1{,}024$ labeled target frames for reference only.}
\end{table*}

\subsubsection{Baseline Failure Mechanisms vs.\ PhyCal}
Table~\ref{tab:baselines_main} confirms all three I/Q failure modes from Section~\ref{sec:related_work} on frozen ResNet-18-Adapt.
Under steep SNR drops, WCP density ratios suffer $\ESS$ collapse ($1{,}024\to{\sim}81$ cross-domain mean, $<40$ on RML22 H$\to$L), leaving WCP, Feature-DR, and ECP under-covered.
Low-SNR score compression breaks OT-CP ($>40$~pp MAE); linear I/Q Mixup in PseudoCal injects unphysical interference inflating temperatures $4\text{--}11\times$ (${\sim}98\%$ over-coverage); and noisy pseudo-labels cause SFCP-Ind to over-cover RML22 H$\to$L.
By contrast, full-$\ESS$ descriptor transport lets zero-label PhyCal-Full-Piv reach 0.84~pp 4C-MAE ($>3\times$ below best baseline QTC at 2.79~pp and below 1{,}024-label Target-LAC* at 0.93~pp), holding all eight scenarios within $[88.9\%, 91.3\%]$ with 76.0\% worst-class coverage (vs.\ Target-APS* 74.4\%), while 0.3\%-MAC PhyCal-Lite-SNR stays at 0.97~pp.
Reflecting the compactness--fairness trade-off, PhyCal-Full-Piv uses a slightly larger 4C-ASS (2.33 vs.\ Target-LAC* 1.82) to cut per-SNR spread $5.8\times$ and boost worst-class coverage, whereas PhyCal-Full-Marg matches Target-LAC* compactness (1.86 ASS).

\begin{figure}[!t]
    \centering
    \includegraphics[width=\columnwidth]{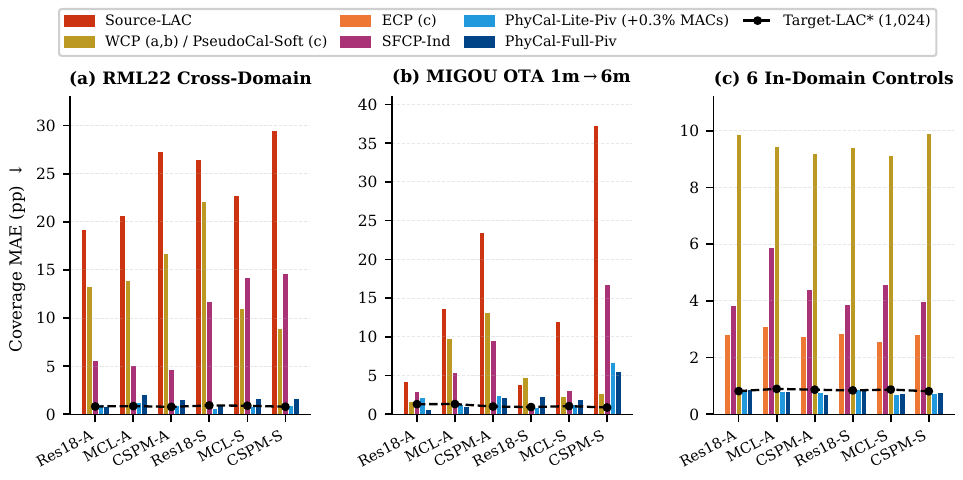}
    \caption{90\% calibration MAE across the $3\times 2$ frozen matrix (Res18-A/S, MCL-A/S, and CSPM-A/S denote adapted/source ResNet-18, MCLDNN, and CSPMNet) using fixed PhyCal-Lite-Piv ($+0.3\%$ MACs) and PhyCal-Full-Piv on (a)~RML22, (b)~USRP OTA MIGOU $1\,\text{m}\to 6\,\text{m}$, and (c)~six in-domain controls across all three benchmarks (Table~\ref{tab:matrix_3x2}).}
    \label{fig:3x2}
\end{figure}

\begin{table*}[!t]
\centering
\caption{Parallel $3\times 2$ Cross-Architecture Matrix: Coverage MAE (pp$\downarrow$) on RML22 (90\%/99\%), MIGOU $1\,\text{m}\to 6\,\text{m}$, Four-Shift Cross-Domain Aggregate (4C, RML22 + MIGOU-MOD), and 6 In-Domain Controls (All 3 Benchmarks, 90\%)}
\label{tab:matrix_3x2}
\renewcommand{\arraystretch}{0.93}
\setlength{\tabcolsep}{3.7pt}
\resizebox{\textwidth}{!}{%
\begin{tabular}{@{}lccccccccccccccc@{}}
\toprule
\multirow{2}{*}{\textbf{Frozen Classifier}} & \multicolumn{4}{c}{\textbf{RML22 Cross (90\% MAE)}} & \multicolumn{3}{c}{\textbf{RML22 Cross (99\%)}} & \multicolumn{3}{c}{\textbf{MIGOU $1\,\text{m}\!\to\!6\,\text{m}$ (90\%)}} & \multicolumn{3}{c}{\textbf{4C Agg.\ (90\% MAE)}} & \multicolumn{2}{c}{\textbf{6 In-Dom.\ (90\%)}} \\
\cmidrule(lr){2-5} \cmidrule(lr){6-8} \cmidrule(lr){9-11} \cmidrule(lr){12-14} \cmidrule(lr){15-16}
 & Src-LAC & SFCP & \textbf{Lite-Piv} & \textbf{Full-Piv} & Src-LAC & \textbf{Lite-Piv} & \textbf{Full-Piv} & Src-LAC & \textbf{Lite-Piv} & \textbf{Full-Piv} & Best Base.$^\dagger$ & \textbf{Lite-Piv} & \textbf{Full-Piv} & \textbf{Cov(\%)} & \textbf{MAE} (Tgt*) \\
\midrule
ResNet-18-Adapt & 19.20 & 5.65 & 1.02 & \textbf{0.83} & 6.76 & \textbf{0.31} & 0.45 & 4.21 & 2.21 & \textbf{0.61} & 2.79 & 1.28 & \textbf{0.84} & 90.06 & 0.89 (0.82) \\
MCLDNN-Adapt    & 20.66 & 5.09 & \textbf{1.21} & 2.05 & 8.86 & \textbf{0.47} & 0.53 & 13.63 & 1.33 & \textbf{0.98} & 5.05 & \textbf{2.97} & 3.48 & 90.21 & \textbf{0.82} (0.90) \\
CSPMNet-Adapt   & 27.34 & 4.71 & \textbf{0.94} & 1.54 & 12.35 & \textbf{0.39} & 0.54 & 23.51 & 2.51 & \textbf{2.22} & 6.40 & \textbf{3.83} & 4.03 & 90.07 & \textbf{0.71} (0.87) \\
\midrule
ResNet-18-Src   & 26.50 & 11.78 & \textbf{0.58} & 0.95 & 19.58 & \textbf{0.45} & 0.50 & 3.86 & \textbf{0.97} & 2.34 & 7.19 & \textbf{2.33} & 3.52 & 90.25 & 0.88 (0.84) \\
MCLDNN-Src      & 22.75 & 14.23 & \textbf{0.86} & 1.68 & 16.92 & \textbf{0.41} & 0.46 & 12.03 & \textbf{1.32} & 1.97 & 8.43 & \textbf{3.33} & 4.58 & 90.36 & \textbf{0.75} (0.87) \\
CSPMNet-Src     & 29.53 & 14.67 & \textbf{0.94} & 1.68 & 17.49 & \textbf{0.48} & 0.55 & 37.33 & 6.73 & \textbf{5.53} & 7.01 & \textbf{5.22} & 5.46 & 90.48 & \textbf{0.78} (0.81) \\
\bottomrule
\end{tabular}}
\vspace{1pt}
\parbox{\textwidth}{\scriptsize Src-LAC: Source-LAC; SFCP: SFCP-Ind; Lite-Piv ($+0.3\%$ MACs, fixed single-pass default matching Fig.~\ref{fig:3x2}) and Full-Piv omit PhyCal-; Cross averages both cross-domain shifts; 4C Agg.\ spans RML22 + MIGOU-MOD; 6 In-Dom.\ spans all three benchmarks (Full-Piv vs.\ Target-LAC* [Tgt*, $N_U{=}1{,}024$] in parentheses). Deterministic Lite-SNR independently reaches $0.58\text{--}1.07$~pp on RML22 90\%, $0.97$~pp on ResNet-18-Adapt 4C (Table~\ref{tab:complexity_lite}), and $1.19$~pp on CSPMNet-Adapt MIGOU $1\,\text{m}\to 6\,\text{m}$. $^\dagger$Best of 12 unlabeled baselines per model (in row order: QTC, Source-APS, QTC, SFCP-Ind, PseudoCal-TopP, WCP).}
\end{table*}

\subsubsection{Cross-Architecture Generalization and Stationary Safety}
Across all six $3\times 2$ models (Table~\ref{tab:matrix_3x2}, Fig.~\ref{fig:3x2}), fixed PhyCal-Lite-Piv and PhyCal-Full-Piv outperform all 12 unlabeled baselines on RML22 ($90\%/95\%/99\%$) and 4C-MAE, cutting error $25\%\text{--}70\%$ over each model's strongest baseline (residual MCLDNN/CSPMNet error concentrates in reverse near-field MIGOU $6\,\text{m}\to 1\,\text{m}$, per Proposition~\ref{prop:dr_decomp}(b)).
Single-pass PhyCal-Lite-Piv holds 90\% RML22 MAE at $0.58\text{--}1.21$~pp across all six models (restoring H$\to$L from $40\%\text{--}57\%$ to $89.3\%\text{--}90.0\%$) and reduces $12\text{--}24$~pp Source-LAC errors on MIGOU $1\,\text{m}\to 6\,\text{m}$ for MCLDNN and CSPMNet-Adapt to $1.32\text{--}2.51$~pp ($0.98\text{--}2.22$~pp for Full-Piv).
Meanwhile, the degradation gate locks all six in-domain controls at $90.1\%\text{--}90.5\%$ (outperforming Target-LAC* on four of six models), while ECP, SFCP-Ind, and PseudoCal-Soft inflate in-domain coverage by $2.6\text{--}9.9$~pp (in-domain MAE in Fig.~\ref{fig:3x2}(c)).

% -----------------------------------------------------------------------------
\subsection{Multi-Risk Reliability, Conditional Fairness, and High-Reliability Hypothesis Reduction}
\label{sec:exp_multialpha}

\begin{figure}[!t]
    \centering
    \includegraphics[width=\columnwidth]{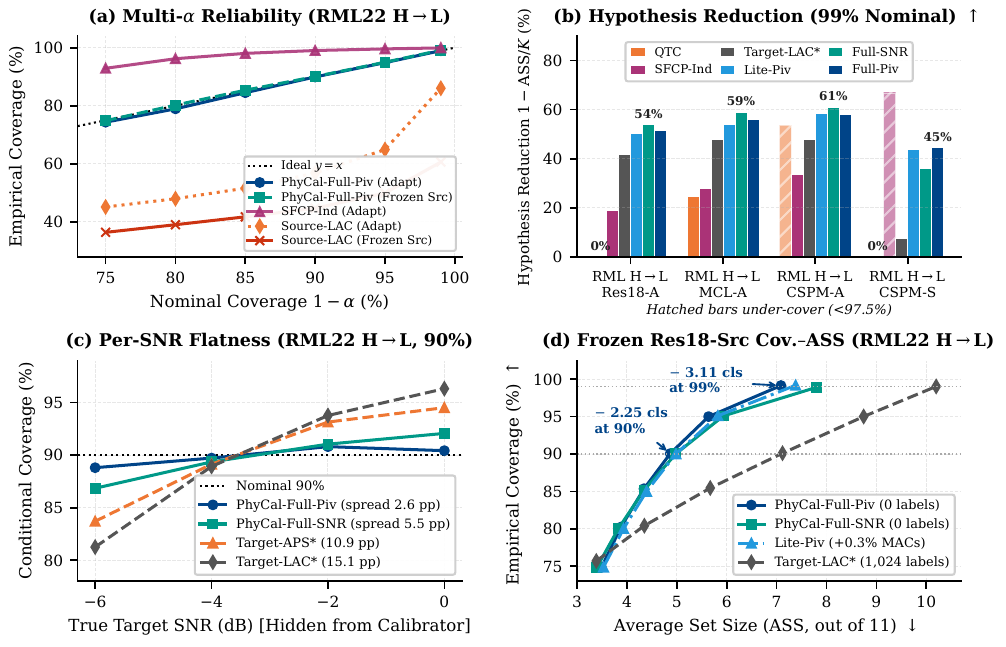}
    \caption{Multi-$\alpha$ reliability and receiver efficiency (model abbreviations follow Fig.~\ref{fig:3x2}): (a)~empirical vs.\ nominal coverage ($75\%\text{--}99\%$, RML22 H$\to$L); (b)~candidate-hypothesis reduction proxy $1-\ASS/K$ at 99\% nominal coverage (hatched bars under-cover $<97.5\%$); (c)~SNR-conditional coverage across unobserved target SNRs (90\%, ResNet-18-Adapt); and (d)~coverage--ASS operating curves on frozen ResNet-18-Src.}
    \label{fig:multialpha}
\end{figure}

\begin{table*}[!t]
\centering
\caption{Multi-$\alpha$ Reliability on ResNet-18-Adapt (RML22 + MIGOU-MOD, Left) and Real 2.4~GHz Multipath OTA 90\% Ranking on Frozen ResNet-18-Adapt ($L = 1{,}024$, Right)}
\label{tab:multialpha_ota}
\renewcommand{\arraystretch}{0.93}
\setlength{\tabcolsep}{3.8pt}
\resizebox{\textwidth}{!}{%
\begin{tabular}{@{}ccccccccc@{}}
\toprule
\multicolumn{6}{c}{\textbf{(A) Multi-$\alpha$ 4C-MAE (pp$\downarrow$) \& RML22 H$\to$L Spread (ResNet-18-Adapt)}} & \multicolumn{3}{c}{\textbf{(B) Real 2.4~GHz Multipath OTA (90\%, ResNet-18-Adapt)}} \\
\cmidrule(lr){1-6} \cmidrule(lr){7-9}
$1-\alpha$ & Src-LAC & SFCP & \textbf{Full-SNR} & \textbf{Full-Piv} (Spread) & Tgt-LAC* (Spread) & \textbf{Method (Frozen ResNet-18-Adapt)} & \textbf{All 4 MAE / Sav.$^\S$} & \textbf{Cross MAE / Sav.$^\S$ (Spr.)} \\
\midrule
75\% & 11.27 & 7.59 & 2.13 & \textbf{1.26} (4.45) & 1.29 (25.17) & \textbf{\#1 PhyCal-Lite-SNR (0.3\%)} & \textbf{3.35} / 61.5\% & 5.89 / 59.4\% (\textbf{6.05}) \\
80\% & 12.16 & 7.45 & 1.91 & \textbf{1.20} (4.05) & 1.23 (22.43) & \textbf{\#2 PhyCal-ZeroProbe-SNR} & \underline{3.58} / 65.7\% & 6.28 / 63.6\% (\underline{6.73}) \\
85\% & 12.81 & 6.65 & 1.75 & \textbf{0.94} (3.63) & 1.00 (19.17) & \textbf{\#3 PhyCal-Full-SNR} & 3.66 / 66.0\% & 6.40 / 64.1\% (6.88) \\
90\% & 10.94 & 4.71 & 1.41 & \textbf{0.84} (2.60) & 0.93 (15.07) & SFCP-Ind$^\ddagger$; Source-APS & 4.09; 4.90 & 2.71/66.6\%; 9.04/72.2\% \\
95\% & 9.35  & 2.63 & 1.29 & \textbf{1.02} (1.81) & 0.66 (8.79)  & QTC; WCP; Source-LAC & 5.46; 5.81; 6.63 & 10.11; 10.70; 12.84 \\
99\% & 3.67  & 0.70 & 0.59 & \textbf{0.37} (0.77) & 0.29 (2.34)  & ECP; PseudoCal-Soft$^\ddagger$ & 5.97; 7.91 & 7.72/73.8\%; 5.84/37.8\% \\
\cmidrule(lr){7-9}
\multicolumn{6}{@{}l@{}}{\scriptsize Spread: RML22 H$\to$L SNR-conditional coverage spread (pp) without true SNR.} & Target-LAC*; APS* ($N_U{=}1{,}024$) & 0.69; 1.00 & 0.69/64.2\%; 0.82/62.9\% \\
\bottomrule
\end{tabular}}
\vspace{1pt}
\parbox{\textwidth}{\scriptsize Src-LAC: Source-LAC; SFCP: SFCP-Ind; Tgt-LAC*: Target-LAC* ($N_U{=}1{,}024$). In (A), 4C spans RML22 + MIGOU-MOD cross-domain shifts and Full- omits PhyCal-. In (B), All 4 / Cross span all four / two cross-domain Multipath OTA pairings; bold/underlined mark top-2 unlabeled All-4 MAE and Cross Spr.; $^\S$Sav.\ reports the raw hypothesis-reduction proxy $1-\ASS/K$~\eqref{eq:demod_saving} for reference (top-3 SNR Cross coverages are $83.7\%\text{--}84.2\%$ under the long-delay boundary); Spr.: per-SNR spread (pp; Target-LAC* 9.39/9.00~pp All-4/Cross); (0.3\%): $+0.3\%$ MACs; semicolon rows report MAE (All 4) and MAE or MAE/Sav.\ (Cross). $^\ddagger$Lower Cross MAE stems from over-conservatism inflating stationary in-domain coverage to 95.5\% (SFCP-Ind; Cross 88.6\%) and 100.0\% (PseudoCal-Soft; All-4 97.8\%).}
\end{table*}

\subsubsection{Multi-$\alpha$ Tracking}
Across $75\%\text{--}99\%$ targets (Table~\ref{tab:multialpha_ota}(A), Fig.~\ref{fig:multialpha}(a)), zero-label PhyCal-Full-Piv closely tracks $y = x$, slightly outperforming 1{,}024-label Target-LAC* across $75\%\text{--}90\%$ on ResNet-18-Adapt and restoring frozen source models from $26\%\text{--}65\%$ under-coverage to within 0.6~pp.

\subsubsection{SNR-Conditional Uniformity and Hard-Class Protection}
Even with 1{,}024 target labels, marginal Target-LAC* leaves a 15.1~pp SNR-conditional coverage spread on RML22 H$\to$L (81.3\% at $-6$~dB vs.\ 96.3\% at $0$~dB; Target-APS* 10.9~pp; Fig.~\ref{fig:multialpha}(c)).
Descriptor-conditioned PhyCal-Full-Piv compresses this spread $5\text{--}7\times$ down to $2.4\text{--}3.1$~pp across all six frozen models (2.6~pp on ResNet-18-Adapt; 2.3~pp for Lite-Piv on MCLDNN-Adapt; 5.5~pp for deterministic PhyCal-Full-SNR), lifts MIGOU $1\,\text{m}\to 6\,\text{m}$ 16-QAM coverage from 58.3\% (Target-LAC*) to 80.4\%, and improves frozen-source coverage--ASS curves, saving 2.25/3.11 classes/frame at 90\%/99\% on ResNet-18-Src (Fig.~\ref{fig:multialpha}(d)) and up to 4.09 classes/frame at 99\% on CSPMNet-Src vs.\ Target-LAC*.

\subsubsection{99\% High-Reliability Savings and Real 2.4~GHz Multipath OTA}
At 99\% target coverage on RML22 H$\to$L, QTC collapses to the full 11-class set (0\% saved) on ResNet-18-Adapt and all three source models (while under-covering CSPMNet-Adapt at 93.3\%), and SFCP-Ind saves at most $19\%\text{--}34\%$ on adapted models while under-covering source models ($83\%\text{--}89\%$).
PhyCal maintains $98.7\%\text{--}99.2\%$ coverage across all six models while cutting candidate-hypothesis workload by $52\%\text{--}61\%$ on adapted backbones (61.1\% on CSPMNet-Adapt, ASS 4.27/11; 59.0\% on MCLDNN-Adapt vs.\ 48.0\%/47.8\% for Target-LAC*) and $36\%\text{--}45\%$ on source backbones vs.\ $7\%\text{--}14\%$ for Target-LAC* (Fig.~\ref{fig:multialpha}(b)).
On native $L = 1{,}024$ HackRF One Multipath OTA without retuning (Table~\ref{tab:multialpha_ota}(B)), PhyCal-Lite-SNR, ZeroProbe-SNR, and Full-SNR rank 1st--3rd among 24 unlabeled methods on ResNet-18-Adapt All-4 90\% MAE (Lite-SNR also ranks 1st at 95\%, 3.63~pp), cut per-SNR spread to 5.95/6.05~pp (All-4/Cross vs.\ Target-LAC* 9.39/9.00~pp), and trim Source-LAC Cross MAE by $6.2\text{--}10.7$~pp across all six models while staying centered in-domain (whereas SFCP-Ind and PseudoCal-Soft reach 2.71/5.84~pp Cross MAE only by inflating stationary in-domain coverage to 95.5\%/100.0\%, worsening All-4 MAE to 4.09/7.91~pp).
PhyCal-SNR outperforms Piv/Marg under equal-SNR long-delay echoes because, even when $\Delta_{\mathrm{snr}}\approx 0$ locks EM on the clean view, continuous quantile regression $\hat{b}_{1-\alpha}(\zeta)$ still adapts per-frame thresholds from spectral and phase descriptors.

% -----------------------------------------------------------------------------
\subsection{Edge Efficiency and Deployment Stress Robustness}
\label{sec:exp_edge}

\begin{figure}[!t]
    \centering
    \includegraphics[width=\columnwidth]{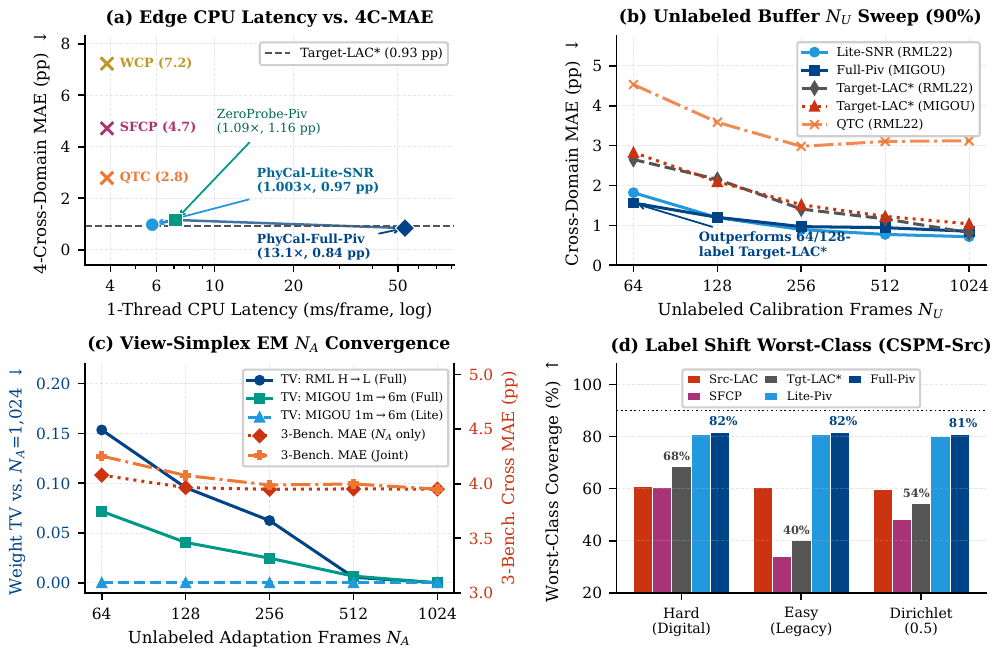}
    \caption{Edge deployability and stress tests: (a)~single-thread Intel Core i7-7800X CPU latency vs.\ 90\% 4C-MAE (ResNet-18-Adapt; PhyCal-Lite-SNR 0.97, ZeroProbe-Piv 1.16, PhyCal-Full-Piv 0.84~pp; baseline $x$-positions use single-pass backbone latency, 3.88~ms; full Piv/SNR breakdown in Table~\ref{tab:complexity_lite}); (b)~buffer $N_U$ sweep; (c)~adaptation buffer $N_A$ sweep (EM TV distance and three-benchmark MAE); and (d)~label-shift worst-class coverage (90\%, CSPMNet-Src [CSPM-Src]; Hard/Easy allocate 70\%/75\% traffic to digital/legacy classes on RML22+MIGOU-MOD / Multipath OTA).}
    \label{fig:edge_robustness}
\end{figure}

\begin{table*}[!t]
\centering
\caption{Hardware Complexity and Latency Profile (RTX 2080 Ti GPU $B = 256$ and Single-Thread Intel Core i7-7800X CPU $B = 1$) Across PhyCal Configurations}
\label{tab:complexity_lite}
\renewcommand{\arraystretch}{0.93}
\setlength{\tabcolsep}{3.8pt}
\resizebox{\textwidth}{!}{%
\begin{tabular}{@{}lccccccccc@{}}
\toprule
\multirow{2}{*}{\textbf{PhyCal Configuration}} & \textbf{Backbone} & \textbf{Extra} & \textbf{MACs Ratio} & \textbf{GPU ms/fr} & \textbf{CPU 1T ms/fr} & \multicolumn{2}{c}{\textbf{Res18-Adapt 90\%}} & \textbf{Res18-Src 90\%} & \textbf{Res18-Adapt 99\%} \\
\cmidrule(lr){7-8}
 & \textbf{Passes} & \textbf{Params} & (Res18 / CSPM) & (Res18 / CSPM) & (Res18 / CSPM) & \textbf{4C-MAE (Piv / SNR)} & \textbf{MIGOU $1\,\text{m}\!\to\!6\,\text{m}$ (Piv)} & \textbf{4C-MAE (Piv)} & \textbf{4C-MAE (Piv)} \\
\midrule
Backbone Only (Source-LAC) & $1\times$ & 0 & $1.000\times$ / $1.000\times$ & 0.015 / 0.016 & 3.88 / 15.30 & 10.94 & 85.8\% (4.21) & 17.58 & 3.67 \\
\textbf{PhyCal-Lite} (9-D $\Zsub$) & \textbf{$1\times$} & \textbf{70.1K} & \textbf{$1.003\times$ / $1.003\times$} & \textbf{0.060 / 0.061} & \textbf{5.78 / 17.20} & 1.28 / \textbf{0.97} & 87.8\% (2.21) & \textbf{2.33} & \textbf{0.36} \\
\textbf{PhyCal-ZeroProbe} (15-D $\Zsig$) & \textbf{$1\times$} & 88.0K & $1.090\times$ / $1.079\times$ & 0.072 / 0.073 & 7.09 / 18.51 & \underline{1.16} / 1.49 & \underline{90.5\% (0.77)} & \underline{2.93} & 0.40 \\
\textbf{PhyCal-Full} (53-D Full) & $13\times$ & 94.1K & $13.09\times$ / $13.08\times$ & 0.403 / 0.472 & 53.24 / 201.70 & \textbf{0.84} / \underline{1.41} & \textbf{89.6\% (0.61)} & 3.52 & \underline{0.37} \\
\bottomrule
\end{tabular}}
\vspace{1pt}
\parbox{\textwidth}{\scriptsize Res18 / CSPM denote ResNet-18 / CSPMNet. Offline View-Simplex EM ($1{,}024\times 8$) takes 10.44~ms on CPU ($B$: batch size; 1T: single-thread Core i7-7800X @ 3.50~GHz; ms/fr: ms/frame). Latencies conservatively sum the frozen backbone pass (omitting the 0.068~ms \emph{-Adapt} quality head), descriptor extraction, and all three 5-fold MLPs in eager PyTorch (PhyCal-Marg adds zero online overhead once $\hat{\tau}_\alpha$ is set; PhyCal-SNR omits rank/view MLPs). 4C-MAE spans RML22 + MIGOU-MOD; MIGOU $1\,\text{m}\to 6\,\text{m}$ gives Cov.\ (MAE, pp) under Piv; PhyCal-Lite runs $9.2\text{--}11.7\times$ faster on 1T CPU than PhyCal-Full.}
\end{table*}

\subsubsection{Hardware Latency and Configuration Trade-Offs}
In Table~\ref{tab:complexity_lite} and Fig.~\ref{fig:edge_robustness}(a), single-pass PhyCal-Lite omits active probes to add just $0.3\%$ MACs ($9.2\text{--}11.7\times$ faster on single-thread CPU than PhyCal-Full), holding ResNet-18-Adapt 4C-MAE at 0.97~pp (SNR) / 1.28~pp (Piv) and outperforming PhyCal-Full on source models (2.33 vs.\ 3.52~pp on ResNet-18-Src).
Its $+1.90$~ms/frame CPU overhead ($+49\%$ on ResNet-18, $+12\%$ on CSPMNet; $0.060\text{--}0.061$~ms/frame GPU) conservatively sums $8\times 8$ Toeplitz/FFT extraction ($0.46$~ms) and eager Python dispatch over all 5-fold small MLPs ($1.44$~ms); PhyCal-Marg adds zero online overhead once $\hat{\tau}_\alpha$ is set, and PhyCal-SNR omits rank/view MLPs ($+0.66$~ms CPU).
Adding only the 15.0K-parameter blind-spot denoiser (PhyCal-ZeroProbe, $+7.9\%\text{--}9.0\%$ MACs) lifts MIGOU $1\,\text{m}\to 6\,\text{m}$ coverage from 87.8\% to 90.5\%, while PhyCal-Full runs in $<0.5$~ms/frame on GPU with a 10.44~ms CPU EM step.

\subsubsection{Small-Buffer and Label-Shift Resilience}
Across all three benchmarks, stress tests (Fig.~\ref{fig:edge_robustness}(b)--(d)) probe finite coherence times and skewed spectrum traffic:
\begin{itemize}
    \item \emph{Small $N_U \in [64, 1{,}024]$}: Combining smooth OOF predictions on $U$ with 1{,}024 source residuals in~\eqref{eq:pbdr_master} halves batch variance and outperforms 64-/128-label Target-LAC* on RML22 and MIGOU-MOD.
    At 99\% coverage, split CP collapses to full sets at $N_U = 64$ ($\lceil 65\times 0.99\rceil > 64$), and PhyCal also remains conservative at $N_U = 64$ ($88\%\text{--}91\%$ full-set rate on ResNet-18-Adapt; hence $N_U \ge 128$ is recommended for 99\% operation); by $N_U = 128$, PhyCal-Lite-Piv on ResNet-18-Adapt saves 58.8\% workload on MIGOU-MOD (0.71~pp MAE, 0.1\% full-set rate $\Prob\{|\calC_\alpha(X)| = K\}$) and 46.7\% on RML22 (0.75~pp, 9.6\% full-set rate).
    \item \emph{Small $N_A \in [64, 1{,}024]$}: With frozen backbones and rank models (isolating EM re-tracking), the gate stays 100\% invariant down to $N_A = 64$ on RML22 and MIGOU-MOD; cutting $N_A$ $8\times$ ($1{,}024\to 128$) shifts three-benchmark ResNet-18-Adapt MAE by only 0.014~pp (Full-Piv) and 0.029~pp (Lite-SNR).
    \item \emph{Skewed class priors}: Against balanced $C_p$, we skew target traffic under three regimes (Fig.~\ref{fig:edge_robustness}(d)): high-order digital (\emph{Hard}: 70\% QPSK, 16-QAM, and 64-QAM on RML22 and MIGOU-MOD; 75\% OFDM, QPSK, and 16-QAM on Multipath OTA), legacy/low-order (\emph{Easy}: 70\% BPSK, GFSK, and WBFM, or 75\% BPSK, GMSK, and WBFM), and symmetric $\mathrm{Dirichlet}(0.5)$.
    PhyCal ranks 1st--3rd among 14 unlabeled methods on ResNet-18-Adapt (Lite-SNR 2.88, Full-SNR 3.30, Full-Piv 3.95~pp).
    Under \emph{Easy} pure label shift, marginal Target-LAC* drops minority high-order worst-class coverage to 40.0\% on CSPMNet-Src (4.9\% on MIGOU-MOD) and 64.2\% on ResNet-18-Adapt, whereas zero-label Full-Piv/Lite-Piv preserve $80.9\%\text{--}81.5\%$ ($77.5\%\text{--}78.3\%$ on MIGOU-MOD; Fig.~\ref{fig:edge_robustness}(d)) and $79.5\%\text{--}79.8\%$, respectively.
\end{itemize}

% -----------------------------------------------------------------------------
\subsection{Physical Mechanism Validation, Ablations, and Configuration Trade-Offs}
\label{sec:exp_physics}

\begin{table*}[!t]
\centering
\caption{Mechanism Validation: Unlabeled View-Simplex EM Channel Discovery (Left) and Paired Stepwise Ablation on ResNet-18-Adapt (RML22 + MIGOU-MOD, 90\%, 3 Seeds, Right)}
\label{tab:ablation_physics}
\renewcommand{\arraystretch}{0.93}
\setlength{\tabcolsep}{3.8pt}
\resizebox{\textwidth}{!}{%
\begin{tabular}{@{}lcccc@{\hspace{10pt}}lccc@{}}
\toprule
\multicolumn{5}{c}{\textbf{(A) Unlabeled View-Simplex EM Channel Diagnosis (PhyCal-Full)}} & \multicolumn{4}{c}{\textbf{(B) Paired Stepwise Ablation (90\% MAE, pp$\downarrow$)}} \\
\cmidrule(lr){1-5} \cmidrule(lr){6-9}
\textbf{Target Scenario} & $\Delta_{\mathrm{snr}}$ (SD) & \textbf{Gate} & \textbf{Discovered $\hat{\pi}$ (Clean / AWGN / Fad.)} & $\ESS_{\mathrm{base}}$ & \textbf{Stepwise Configuration} & \textbf{4C-MAE (SD)} & \textbf{8S-MAE} & \textbf{MIGOU $1\,\text{m}\!\to\!6\,\text{m}$} \\
\cmidrule(lr){1-5} \cmidrule(lr){6-9}
In-Domain (4 scenarios) & $0.00\text{ to }+0.03$ & Clean & $100\% / 0\% / 0\%$ & 1{,}024 & \textit{M1--M2 Rank-Model Ablation (Fixed Clean-View Res.):} & & & \\
RML22 L$\to$H & $+2.84$ & Clean & $100\% / 0\% / 0\%$ & 1{,}024 & (1) 5-Fold Baseline (15-D classical statistics) & 2.06 (0.33) & 1.62 & 86.0\% (3.95) \\
MIGOU $6\,\text{m}\to 1\,\text{m}$ & $+3.06$ & Clean & $100\% / 0\% / 0\%$ & 1{,}024 & (2) + 8-View Suite + Tail-W.\ RPS + Noise Mono. & 1.89 (0.33) & 1.53 & 87.0\% (2.96) \\
\textbf{RML22 H$\to$L} & \textbf{$-1.66$} & \textbf{EM} & \textbf{$0.0\% / \mathbf{100.0\%} / 0.0\%$} & \textbf{1{,}024} & (3) + Toeplitz + $U(1)$ + AWGN Probes & 1.68 (0.21) & 1.44 & 87.0\% (2.98) \\
\textbf{MIGOU $1\,\text{m}\to 6\,\text{m}$} & \textbf{$-1.39$} & \textbf{EM} & \textbf{$85.0\% / 0.0\% / \mathbf{15.0\%}$} & \textbf{1{,}024} & (4) + Blind-Spot Denoiser (PhyCal-Full) & \textbf{1.12 (0.17)} & \textbf{1.15} & \textbf{89.9\% (0.85)} \\
\cmidrule(lr){1-5} \cmidrule(lr){6-9}
\textit{Control: 15-D Classical Statistics} & & & & & \textit{M3--M4 Transport \& Mode Ablation (PhyCal-Full):} & & & \\
MIGOU $1\,\text{m}\to 6\,\text{m}$ & $-0.13$ & Clean$^\dagger$ & Blind to OTA shift & 1{,}024 & Pooled-View Binary DR (Marginal) & 2.80 & 2.48 & 92.0\% (2.06) \\
MIGOU $6\,\text{m}\to 1\,\text{m}$ & $+0.12$ & Clean & Blind to OTA shift & 1{,}024 & View-Simplex DR (PhyCal-Full-Marg) & 1.55 & 1.42 & 92.2\% (2.25) \\
\textit{Synthetic-View OOF Check} & \multicolumn{4}{l}{Clean 90.4\%, AWGN ($\eta{=}1$) 91.5\%, Mild Fad.\ 89.0\% (vs.\ Src-LAC 21.7--67.8\%)} & View-Simplex + Pivotal (PhyCal-Full-Piv) & \textbf{0.84 (0.30)} & \textbf{0.90} & \textbf{89.6\% (0.61)} \\
\bottomrule
\end{tabular}}
\vspace{1pt}
\parbox{\textwidth}{\scriptsize In (A), AWGN / Fad.\ sum $\hat{\pi}$ over $v{=}1\text{--}5$ / $v{=}6,7$; Mild Fad.: $v{=}6$; Src-LAC: Source-LAC. In (B), the top block fixes the clean-view marginal OOF residual (no EM) to isolate Module~1--2 gains, while the bottom block compares Module~3--4 transport/modes on the same PhyCal-Full rank model; Tail-W.\ RPS: tail-weighted RPS; Noise Mono.: penalty~\eqref{eq:bw_loss}; MIGOU $1\,\text{m}\to 6\,\text{m}$ gives Cov.\ (MAE, pp). $^\dagger$Classical 15-D statistics yield only $\Delta_{\mathrm{snr}}{=}{-}0.13$~SD on MIGOU $1\,\text{m}\to 6\,\text{m}$; dropping them from PhyCal-Full slightly improves 90\%/99\% Piv MAE on CSPMNet-Src ($5.46/2.29\to 5.37/2.25$~pp) and 99\% ResNet-18-Adapt ($0.37\to 0.34$~pp).}
\end{table*}

\subsubsection{Blind Channel Discovery and Paired Ablations}
Table~\ref{tab:ablation_physics}(A) confirms faithful blind channel identification at $\ESS_{\mathrm{base}} = 1{,}024$: the gate locks onto the clean view in all stationary and upward-SNR links, while EM discovers 100.0\% AWGN on RML22 H$\to$L and an 85\% clean / 15\% Rician-CFO mix (0\% AWGN) on MIGOU $1\,\text{m}\to 6\,\text{m}$---a shift missed by 15-D classical statistics ($\Delta_{\mathrm{snr}} = -0.13$~SD).
In paired 5-fold rank-model ablations with fixed clean-view OOF residuals (Table~\ref{tab:ablation_physics}(B), top), adding the 8-view suite with noise monotonicity, subspace/active probes, and blind-spot residuals monotonically reduces 4C-MAE ($2.06\to 1.89\to 1.68\to 1.12$~pp) and halves seed SD.
On the fixed PhyCal-Full rank model (bottom), replacing Pooled-View Binary DR (one pooled binary discriminator against $A$ forming sample weights in~\eqref{eq:pbdr_master}) with View-Simplex EM and OOF RPS recalibration prevents mismatched noisy views from biasing source residuals ($-44.7\%$ marginal MAE, $2.80\to 1.55$~pp), reaching 0.84~pp with pivotal scores; held-out synthetic-view OOF checks maintain ${\sim}90\%$ coverage where Source-LAC falls to $21.7\%\text{--}67.8\%$.

\subsubsection{Why PhyCal-Lite Outperforms PhyCal-Full on Unadapted Classifiers}
In Table~\ref{tab:matrix_3x2}, single-pass PhyCal-Lite-Piv ($+0.3\%$ MACs) outperforms 53-D PhyCal-Full-Piv ($13\times$ passes) in RML22 90\% and 4C-MAE on five of six models---including all three unadapted (\emph{-Src}) classifiers (e.g., 2.33 vs.\ 3.52~pp on ResNet-18-Src; 3.33 vs.\ 4.58~pp on MCLDNN-Src)---and all six on RML22 at 99\%.
This suggests a bias--variance trade-off: when unadapted softmax posteriors distort abruptly under target noise, feeding 23-D active classifier probes ($U(1)$/AWGN queries of $f$) into the compact rank model injects $f$'s representation drift and increases variance vs.\ classifier-free 9-D $\Zsub$.
On smoothly adapted ResNet-18-Adapt, active probes add real stability signal, letting PhyCal-Full-Piv achieve the lowest overall error (0.84~pp 4C-MAE, 2.6~pp spread, 0.61~pp MIGOU $1\,\text{m}\to 6\,\text{m}$), validating Section~\ref{sec:module34}'s selection guide.

\subsubsection{Limitations and Physical Identifiability Boundaries}
Two empirically verified regimes delineate Proposition~\ref{prop:dr_decomp}(b)'s out-of-hull boundaries:
(i)~\emph{Reverse near-field OTA drift} (MIGOU $6\,\text{m}\to 1\,\text{m}$, $\Delta_{\mathrm{snr}} = +3.06$~SD): since the target link is cleaner than the source, the one-way gate locks $\hat{\pi} = e_0$, leaving 90\% coverage at $81.1\%\text{--}81.9\%$ (MCLDNN-Adapt), $78.1\%\text{--}79.2\%$ (CSPMNet-Adapt), and $77.0\%\text{--}79.7\%$ (MCLDNN-Src), while ResNet-18-Adapt stays at $89.1\%\text{--}89.8\%$; its recurrence across both regimes indicates an identifiability boundary of the forward-degradation channel suite under non-AWGN near-field distortion rather than missing adapter capacity.
(ii)~\emph{Equal-SNR long-delay multipath} (Multipath OTA, $[0,100,200]\,\mu\text{s} = [0,200,400]$ samples at 2~MSps, $\Delta_{\mathrm{snr}}\in[-0.12,+0.03]$~SD): echo delays far exceed the 8-lag Toeplitz window and 15-sample blind-spot receptive field (ungated EM still places $99.5\%$ mass on the clean view), leaving $5.89\text{--}6.40$~pp cross-domain 90\% MAE on ResNet-18-Adapt (still $6.2\text{--}10.7$~pp below Source-LAC across all six models).
More broadly, severe nonlinear front-end distortion or cross-transmitter oversampling mismatch (concept shift) falls outside our fixed-pulse-shape AWGN/short-lag Rician-CFO hull, motivating wideband long-lag descriptors and explicit out-of-distribution monitoring.

% =============================================================================
\section{Conclusion}
\label{sec:conclusion}
% =============================================================================

We proposed \textbf{PhyCal}, a strictly post-hoc, classifier-agnostic framework for zero-label conformal calibration under wireless channel shift.
Combining classifier-free baseband descriptors, an eight-view synthetic channel suite, noise-monotonic rank modeling, and degradation-gated View-Simplex EM transport, PhyCal avoids sample collapse to deliver doubly robust prediction sets with improved conditional coverage uniformity without retraining classifiers or labeling target data.
Across synthetic SNR drops, real USRP OTA links, and real 2.4~GHz HackRF One multipath channels over six frozen ResNet-18, MCLDNN, and CSPMNet models, PhyCal outperforms all 12 unlabeled baselines on RML22, MIGOU-MOD, and overall four-scenario multipath reliability, matches 1{,}024-label target calibration over $75\%\text{--}90\%$, compresses per-SNR spread $5\text{--}7\times$, and cuts candidate-hypothesis blind demodulation workload by $52\%\text{--}61\%$ ($36\%\text{--}45\%$ unadapted) at 99\% coverage.
Single-pass PhyCal-Lite ($+0.3\%$ MACs, $9.2\text{--}11.7\times$ single-thread CPU speedup) retains near-full accuracy---outperforming PhyCal-Full on unadapted models---while resisting small buffers and skewed priors.
Future work will extend $\Zsub$ to wideband long-delay multipath, RF front-end drift, and open-set out-of-distribution monitoring.

\appendices

\section{Proof of Proposition~1 (EM Concavity and Maximal ESS)}
\label{app:prop1}
For part~(1) of Proposition~\ref{prop:em_ess}, equal view priors over $v\in\{0,\dots,V-1\}$ imply $\sum_v \pi_v p_{S,v}(\zeta) = V\bar{p}_S(\zeta)\sum_v \pi_v \hat{P}(v\mid\zeta)$.
Dropping $\pi$-independent factors leaves $\mathcal{J}(\pi)$ in Proposition~\ref{prop:em_ess}, an average of compositions of concave $\log(\cdot)$ with positive affine maps on $\Delta^{V-1}$, which EM~\eqref{eq:view_em} monotonically ascends via Jensen's inequality~\cite{saerens2002adjusting,alexandari2020mlls} to a global maximizer.
For part~(2), Cauchy--Schwarz over $i\in\{1,\dots,N_{\mathrm{base}}\}$ gives $\hat{\pi}_v^2 = (\sum_i w_{v,i})^2 \le N_{\mathrm{base}}\sum_i w_{v,i}^2$ and $1 = (\sum_i W_i)^2 \le N_{\mathrm{base}}\sum_i W_i^2$, with equality if and only if $w_{v,i} = \hat{\pi}_v/N_{\mathrm{base}}$ and $W_i = 1/N_{\mathrm{base}}$ for all $i$; uniform weights $w_{v,i}^*$ uniquely achieve both maxima simultaneously.

\section{Proof of Proposition~2 and Remark~1 (Uniform DR Rates and Cross-Fitted Scores)}
\label{app:prop2}
Substituting $F_T = \mu_T + \delta_T$ into $\widetilde{F}_T^{\mathrm{raw}} = \widehat{\mu}_U + \widehat{\delta}_{S,\hat{\pi}}$ and adding/subtracting $\delta_{S,\hat{\pi}}$ yields~\eqref{eq:dr_decomp}.
For Term~(I), conditioned on $(C_p, A)$, the variables $G_{\tilde{q}}(x,t)\in[0,1]$ ($x\in U$) are i.i.d.\ with mean $\mu_T(t)$; because $t\mapsto G_{\tilde{q}}(x,t)$ is pointwise non-decreasing in $[0,1]$ (a totally ordered VC-subgraph class of dimension~1), Hoeffding and empirical-process bounds yield $\sup_{t\in\R}|\widehat{\mu}_U(t) - \mu_T(t)| = O_p(N_U^{-1/2})$.
For Term~(II), although $\hat{\pi}\in\Delta^{V-1}$ depends on $(C_p, A)$, simplex linearity bounds $|\widehat{\delta}_{S,\hat{\pi}}(t) - \delta_{S,\hat{\pi}}(t)| \le \max_{0\le v\le V-1}|\widehat{\delta}_{S,v}(t) - \delta_{S,v}(t)|$.
For each view $v$ and balanced fold $k\in\{1,\dots,5\}$ ($|I_k| = N_{\mathrm{base}}/5$), define $\delta_{S,v}^{(-k)}(t) = \E_{P_{S,v}}[\Ind\{\tilde{s}(X,Y)\le t\} - G_{\tilde{q}}^{(-k)}(X,t)\mid C_p^{(-k)}, A]$.
Conditioned on $(C_p^{(-k)}, A)$, $\tilde{q}^{(-k)}$ is fixed and the held-out indicators $\Ind\{\tilde{s}(x_{v,i}^S, y_i^S)\le t\}$ and monotone maps $G_{\tilde{q}}^{(-k)}(x_{v,i}^S, t)$ ($i\in I_k$) are i.i.d.\ in $[0,1]$ with VC-subgraph dimension~1, so Dvoretzky--Kiefer--Wolfowitz (DKW) and empirical-process bounds with total expectation control each fold--view deviation uniformly in $t$ at rate $O_p(N_{\mathrm{base}}^{-1/2})$.
By linearity of $G_{\tilde{q}}$ in $\tilde{q} = \frac{1}{5}\sum_{k=1}^5 \tilde{q}^{(-k)}$, $\frac{1}{5}\sum_{k=1}^5 \delta_{S,v}^{(-k)}(t) = \delta_{S,v}(t)$ identically; a union bound over all $5V$ fold--view pairs yields $\sup_{t\in\R}|\widehat{\delta}_{S,\hat{\pi}}(t) - \delta_{S,\hat{\pi}}(t)| = O_p(N_{\mathrm{base}}^{-1/2})$ uniformly over $\hat{\pi}\in\Delta^{V-1}$.
In our implementation~\eqref{eq:nested_recal}, Stage-2 stacking fits only $2V{=}16$ scalar parameters on Stage-1 OOF predictions and applies them to $\bar{q}$ on $U\cup T$; strict fold-wise independence and exact linearity hold identically for raw 5-fold OOF rank models or doubly nested inner-OOF recalibration, while low-DOF Stage-2 stacking introduces an asymptotically negligible discrepancy absorbed into $\Delta_{\mathrm{ens}}(t)$.
Under condition~(a), either $\mu_D = F_D$ gives $\delta_T = \delta_{S,\hat{\pi}} = 0$, or $P_T = \sum_v \pi_v^* P_{S,v}$ gives $\delta_T = \sum_v \pi_v^*\delta_{S,v} = \lim_{\hat{\pi}\to\pi^*}\delta_{S,\hat{\pi}}$.
Outside the convex hull of $\{P_{S,v}\}$ under misspecified $\tilde{q}$, $\delta_T$ is unidentifiable without target labels, establishing~(b).

For Remark~\ref{rem:adaptive_scores}, let $\tilde{s}^{(-k)}$ be the fold-$k$ OOF score fitted on $(C_p^{(-k)}, A)$ and set $\bar{\delta}_{S,\hat{\pi}}^{\mathrm{OOF}}(t) = \frac{1}{5}\sum_{v,k}\hat{\pi}_v \delta_{S,v}^{(-k)}(t)$.
Conditioned on $(C_p^{(-k)}, A)$, both $\tilde{s}^{(-k)}$ and $\tilde{q}^{(-k)}$ are fixed, so the same uniform DKW/VC bound gives $\sup_{t\in\R}|\widehat{\delta}_{S,\hat{\pi}}(t) - \bar{\delta}_{S,\hat{\pi}}^{\mathrm{OOF}}(t)| = O_p(N_{\mathrm{base}}^{-1/2})$, leaving $\Delta_{\mathrm{ens}}(t) = \bar{\delta}_{S,\hat{\pi}}^{\mathrm{OOF}}(t) - \delta_{S,\hat{\pi}}(t)$ in~\eqref{eq:dr_decomp}.
Under PhyCal-Piv, for any rank PMF $q$ with cumulative sums $Q_r = \sum_{r'=1}^r q_{r'}$ ($Q_0{=}0$) and $u_x\sim\mathrm{Unif}(0,1)$, integrating over $u_x$ yields $\E_{u_x}[G_q(x,t)] = \sum_{r=1}^K \min(q_r, [t - Q_{r-1}]_+) \equiv t$ for all $t\in[0,1]$ and all $x$; hence $\mu_{S,v}^{(-k)}(t) = \mu_{S,v}(t) = t$ cancels and $\Delta_{\mathrm{ens}}(t) = \frac{1}{5}\sum_{v,k}\hat{\pi}_v(F_{S,v}^{(-k)}(t) - F_{S,v}(t))$ on $[0,1]$.

\section{Proof of Corollary~1 (Isotonic Coverage Control)}
\label{app:cor1}
Since $F_T(t)\in[0,1]$, clipping $\widetilde{F}_T^{\mathrm{raw}}$ to $[0,1]$ cannot increase pointwise distance to $F_T$, so the clipped curve $g$ satisfies $F_T - \varepsilon_n \le g \le F_T + \varepsilon_n$.
Because Pool-Adjacent-Violators isotonic projection $\Pi_{\mathrm{iso}}$ is order-preserving and leaves the non-decreasing shifts $F_T \pm \varepsilon_n$ invariant, we obtain $F_T - \varepsilon_n \le \Pi_{\mathrm{iso}}(g) = \widetilde{F}_T \le F_T + \varepsilon_n$, i.e., $\|\widetilde{F}_T - F_T\|_\infty \le \varepsilon_n$.
By right-continuity of $\widetilde{F}_T$ and~\eqref{eq:threshold_inv}, $\widetilde{F}_T(\hat{\tau}_\alpha) \ge \gamma_n$, which yields~\eqref{eq:cov_lower}: $F_T(\hat{\tau}_\alpha) \ge \widetilde{F}_T(\hat{\tau}_\alpha) - \varepsilon_n \ge \gamma_n - \varepsilon_n$.
When $F_T$ is continuous, taking the left limit gives $F_T(\hat{\tau}_\alpha) = F_T(\hat{\tau}_\alpha^-) \le \widetilde{F}_T(\hat{\tau}_\alpha^-) + \varepsilon_n \le \gamma_n + \varepsilon_n$, hence $|F_T(\hat{\tau}_\alpha) - (1-\alpha)| \le \varepsilon_n + (1-\alpha)/N_U$.

% \section*{Acknowledgments}
% [Placeholder: The authors thank the anonymous reviewers and associate editor for their constructive comments. This work was supported by [Funding Agency] under Grant [Grant Number].]

\bibliographystyle{IEEEtran}
\bibliography{references}

\vfill

\end{document}